\documentclass[
showpacs,
showkeys,
aps, prb,%
amsmath,amssymb,
reprint,%
,longbibliography
]{revtex4-2}
\usepackage{graphicx}
\usepackage{isomath}  
\usepackage{bm}
\graphicspath{{figs/}}
\usepackage[hidelinks,colorlinks=true,linkcolor=blue,citecolor=blue]{hyperref}

\begin{document}


\title[LDOS]{Machine learning the local electronic density of states}

\author{A. Aryanpour}
\affiliation{Department of Physics, Shahid Beheshti University, Tehran, Iran} 
\author{Ali Sadeghi}%
\affiliation{Department of Physics, Shahid Beheshti University,  Tehran, Iran} 

\date{\today}


\begin{abstract}	
Electronic density of states (DOS) plays a crucial role in determining and understanding materials properties.
We investigate the machine learnability of additive atomic contributions to electronic DOS,
   focusing on atom-projected DOS rather than structural DOS. 
  This approach for structure-property mapping is both scalable and transferable, 
  and achieves high prediction accuracy for pure and compound silicon and carbon structures of various sizes and configurations. 
  Furthermore, we demonstrate the generalizability of this model to complex Sn-S-Se compound structures. 
Utilizing locally trained DOS is shown to significantly enhance the accuracy of predicting material properties,
	  including band energy, Fermi energy, heat capacity, and magnetic susceptibility. 
Our findings indicate that directly learning atomic DOS, rather than structural DOS, improves the efficiency, accuracy, and interpretability of machine learning in structure-property mapping. 
	This streamlined approach reduces computational complexity, paving the way for examination of electronic structures in materials without the need for computationally expensive \emph{ab initio} calculations.
	\end{abstract}

\maketitle

\section{\label{sec:intro}Introduction} 
In the last decade, low-cost calculation of the electronic structure of materials has increasingly been taken into consideration for both pure research and applied purposes.  
Compared to quantum chemistry methods which determine the many-body wavefunction,
the density functional theory (DFT) provides a much faster computational scheme 
by solving the single-electron Kohn-Sham equations
for the  ground state electron density~\cite{Kohn1, Kohn2}.
    Although the reliability of DFT calculations has been continuously improved through the decades~\cite{sci16},
 its application is still limited to systems that are only a few orders of magnitude 
larger than those tractable with quantum chemistry methods.  
The computational bottleneck of DFT  
lies in the self-consistent construction of large Hamiltonian matrices, 
which involves repeated diagonalization.
The unknown true form of the exchange and correlation (XC) functionals 
also limits the accuracy of this approach.
To speed up electronic structure calculations and extend its application to larger systems,
one may approximate the DFT Hamiltonian with a tight binding-type Hamiltonian which can 
partially be filled with pre-calculated reference values~\cite{DFTB98}
or trained on {\it ab initio} eigenvalues by machine learning (ML)~\cite{deepH22,MLTB24}.
Alternatively, an ultimate outcome of electronic structure calculations,
either scalar, vectorial or tensorial quantities,
might be directly predicted with ML.
For example, the potential energy surface of an atomic structure 
or the Hellmann-Feynman forces on the atomic centers as required for molecular dynamics simulations 
can be the output layer of an artificial neural network.~\cite{Behler,Behler07,Ko21}
This approach bypasses the costly construction and digitalization of the Hamiltonian matrices
but it also ignores the whole physics of the problem. 

The distribution of electronic energy levels of a system,
known as the density of states (DOS),
is a key quantity in analyzing physical phenomena and designing materials with desired electronic and optical properties.
Although the {\it structural} DOS is a {\it global} quantity that depends on the atomic structure as a whole,
it is a common practice to decompose it into orbital-{\it projected} components, known as PDOS.
Then one can interpret physical properties and chemical behaviour 
in terms of contributions from the constituting atoms or atomic orbitals.
Such an understanding is helpful in designing catalysts, semiconductor devices, or nanostructure materials,
where electronic characteristics can substantially vary at the nanoscale. 
This is also crucial when the functionality of the material critically depends  on heterogeneity at the atomic scale. 

Generally, decomposition of a physical quantity into atomic contributions 
 leads to two important computational advantages: scalability and transferability.
 The atomic contributions are first trained, with low computational effort, on small-sized and basic structures 
 but are then applicable to systems of much larger sizes and complexities.
 The underlying assumption is locality of atomic contribution. 
The intuition behind locality is mostly based on the \emph{nearsightedness principle},
referring to the pioneering work by Kohn and Prodan~\cite{Kohn2,Prodan05, *Prodan06}
who showed that for a many-electron system 
the electron density at a given point is practically insensitive to 
the change of the external potential at large distances. 
In practice, it has also been shown by DFT calculations that 
in the absence of chemically impactful defects like aliovalent impurities or vacancies,
the electronic matter in bulk (but not in low dimensional samples) remains fairly nearsighted.~\cite{Babaei24}
We aim to explore in this study whether the atom-projected DOS, 
referred to as \emph{local DOS} (LDOS) hereafter, depends merely on the local environment of the atoms.

DFT calculations provide DOS (either structural DOS or atomic LDOS as the sum of PDOS over orbitals of a given atom)
as an array of real numbers distributed on the energy interval. 
The array might serve as the input feature vector into a  ML model for predicting physical quantities~\cite{Fung21,Knosgaard22},
topological invariants~\cite{topodos25} and so on.
Alternatively, DOS, as an array, can be the target of ML prediction.~\cite{benmahmoud, adaptivedos25} 
In this study we aim to show the highly employability  of the locality nature of atomic LDOS for a supervised learning
and thus an improved efficiency of learning LDOS instead of structural DOS.
In the rest of this manuscript, we first explain the mathematical background behind local learning of DOS
and then introduce
our ML procedure and  training dataset. 
The prediction error for LDOS and DOS are compared for selected silicon and carbon systems. 
We then show the performance of predication of a few of the derived physical properties from the trained LDOS,
including band energy, Fermi energy and DOS at the Fermi level, magnetic susceptibility and excitation distribution spectrum before we draw our conclusions.

\section{Theoretical Background and Methods}
 
\subsection{Local Density of States}\label{sec:LDOS}  
We denote the total (or \emph{structural}) DOS by $\mathcal{D}$, such that $\mathcal{D}(\varepsilon) \, d\varepsilon$ represents the number of states available for electrons within the energy interval $\varepsilon$ to $\varepsilon + d\varepsilon$ throughout the entire structure. The central idea of this work is that the structural DOS can be split up as 
\begin{equation} \label{eq:sumDi}  
\mathcal{D}(\varepsilon) = \sum_i \mathcal{D}_i(\varepsilon),  
\end{equation}  
to atomic contributions $\mathcal{D}_i(\varepsilon)$ which can be defined in several ways~\cite{onetep19}. 
We emphasize that the method used in Ref.~\onlinecite{benmahmoud} is not one of the \emph{physical} projection schemes explained in the following.
Instead, they machine learned $\mathcal{D}_i$ by minimizing the difference between the total DOS serving as the reference and the predicted sum of atomic $\mathcal{D}_i$, as  subsequently discussed. When necessary, we differentiate between physical projection and machine learning splitting of DOS by referring to them as projected DOS and split DOS, respectively.

For a crystalline system with a Brillouin zone (BZ) of volume $\Omega_\text{BZ}$, the DOS is given by  
\begin{equation} \label{eq:DOS0}  
\mathcal{D}(\varepsilon) =   
\frac{1}{\Omega_\text{BZ}} \sum_n \int_{\text{BZ}}   
\delta \left(\varepsilon - \varepsilon_n(\mathbf{k})\right) \, d\mathbf{k}.  
\end{equation}  
Electronic structure calculations provide eigenfunctions $|\psi_n(\mathbf{k})\rangle$ and corresponding energies $\varepsilon_n(\mathbf{k})$ at a finite number $N_k$ of grid points in ${\mathbf{k}}$-space. Thus, the latter equation is approximated by  
\begin{multline} \label{eq:D2}  
\mathcal{D}(\varepsilon) =  
\frac{1}{N_\mathbf{k}} \sum_{n,{\mathbf{k}}}   
\delta (\varepsilon - \varepsilon_{n,\mathbf{k}})  
\langle \psi_{n\mathbf{k}} | \left( \int |{\bf r}\rangle \langle {\bf r}| \, d{\bf r} \right) |\psi_{n\mathbf{k}} \rangle
\\ =   
\frac{1}{N_\mathbf{k}} \sum_{n,{\mathbf{k}}}   
\int \left| \psi_{n\mathbf{k}} ({\bf r}) \right|^2   
\delta (\varepsilon - \varepsilon_{n,\mathbf{k}}) \, d{\bf r}.  
\end{multline}  
Note that we have employed the normalization of the eigenfunctions $1=\langle\psi_{n\mathbf{k}}|\psi_{n\mathbf{k}}\rangle$ and the completeness relation $1=\int |{\bf r}\rangle \langle {\bf r}| \, d{\bf r}$.   
For a non-periodic sample such as a  molecule or atomic cluster, $N_\mathbf{k} = 1$, and calculations are performed only at the point ${\mathbf{k}}=(0,0,0)$. 

To transform this equation into Eq.~(\ref{eq:sumDi}), the whole space is partitioned into non-overlapping atomic basins, and $\int \rightarrow \sum_i \int_i$, such that the atom-projected LDOS for each atomic basin reads   
\begin{equation} \label{eq:Di_r}  
\mathcal{D}_i(\varepsilon) = \int\limits_{\text{atom }i} \mathcal{D}(\varepsilon, {\bf r}) \, d{\bf r},  
\end{equation}  
where the space-resolved DOS is given by  
\begin{equation} \label{eq:Dr}  
\mathcal{D}(\varepsilon, {\bf r}) =   
\frac{1}{N_\mathbf{k}} \sum_{n,{\mathbf{k}}}   
\left| \psi_{n\mathbf{k}} ({\bf r}) \right|^2   
\delta (\varepsilon - \varepsilon_{n,\mathbf{k}}).  
\end{equation}  
This is a physical quantity which is directly measured by means of a scanning probe microscope~\cite{corral} and interpreted by the Tersoff-Hamann model~\cite{THmodel85}. Notably, $\mathcal{D}(\varepsilon, {\bf r})$ can be machine learned on a real space mesh from the atomic environment surrounding each mesh node~\cite{Chandrasekaran}. In some of the literature~\cite{topodos25}, $\mathcal{D}(\varepsilon, {\bf r})$ is referred to as the {\it local} DOS. But the same term has also been used for the atomic contribution $\mathcal{D}_i(\varepsilon)$ defined by Eq.~(\ref{eq:Di_r}) or (\ref{eq:Di_a}), which is adopted in the present study.  

Alternatively, one may use atom-centered orbitals as the basis set, expressing the wavefunction as $|\psi_{n\mathbf{k}}\rangle = \sum_\alpha c_{{n\mathbf{k}},\alpha}|\phi_{\alpha} \rangle$. By employing the completeness of the basis set, $1=\sum_\alpha |\phi_{\alpha} \rangle \langle \phi_\alpha|$, and the normalization of the eigenfunctions, $1=\langle\psi_{n\mathbf{k}}|\psi_{n\mathbf{k}}\rangle$, we obtain (cf. Eq.~(\ref{eq:D2}))  
\begin{multline} \label{eq:D}  
\mathcal{D}(\varepsilon) =   
\frac{1}{N_\mathbf{k}} \sum_{n,{\mathbf{k}}}   
\delta (\varepsilon - \varepsilon_{n,\mathbf{k}})   
\langle \psi_{n\mathbf{k}} | \left( \sum_\alpha \left |\phi_{\alpha} \rangle \langle \phi_\alpha \right| \right) |\psi_{n\mathbf{k}} \rangle 
\\ = \frac{1}{N_\mathbf{k}} \sum_{n,{\mathbf{k}}} \sum_\alpha   
\left| c_{{n\mathbf{k}},\alpha} \right|^2 \delta (\varepsilon - \varepsilon_{n,\mathbf{k}}).  
\end{multline}  
Thus, the second physically meaningful projection scheme of DOS, as expressed in Eq.~(\ref{eq:sumDi}), is formulated, where  
\begin{equation} \label{eq:Di_a}  
\mathcal{D}_i(\varepsilon) = \frac{1}{N_\mathbf{k}} \sum_{n,{\mathbf{k}}} \sum_{\alpha \in i}   
\left| c_{{n\mathbf{k}},\alpha} \right|^2 \delta (\varepsilon - \varepsilon_{n,\mathbf{k}}),  
\end{equation}  
depends solely on those $\alpha$ components of the eigenvectors that are centered on atom $i$. This equation is commonly utilized to decompose the energy distribution of states into angular momentum components (orbital-resolved PDOS), providing detailed insights into the electronic activity of individual atomic orbitals.  

 \subsection{Additivity}\label{sec:additivity}
One of the earliest, most common applications of ML in materials science involves regenerating the high-dimensional potential energy surface of atomic structures. Here, artificial neural networks (ANN) are trained on a rich and diverse set of structure-energy pairs to predict interatomic forces and the energies of large systems at moderate computational cost.~\cite{Behler07,Ko21} This mapping of structure to energy is a practically important example of the general \emph{structure-property mapping} strategy
\begin{equation} \label{eq:S2P}
{\bf \mathcal{S}} \mapsto {\bf \mathcal{P}}.
\end{equation}
The structure ${\bf \mathcal{S}} \equiv \{(Z_i, {\bf R}_i)\}_{i=1}^N$ represents a collection of $N$ atoms of types $Z_i$ at positions ${\bf R}_i$. The target property ${\bf \mathcal{P}}$ can be one of several forms, including:
a single real number, such as potential energy~\cite{Behler07} or energy band gap~\cite{gap21},
a vector of real numbers, representing properties like DOS~\cite{benmahmoud}, atomic charges~\cite{Babaei20}, or atomic forces~\cite{Behler07,Ko21},
or even a scalar field such as electron density~\cite{Burke17bypass,Ceriotti18,Chandrasekaran} and its functionals~\cite{SCAN,Burke12func}, or space-projected DOS~\cite{Chandrasekaran}.
In any case, the ML model is expected to be \emph{transferable} to complex samples different from those used for training. This feature significantly reduces the computational complexity of the ML structure-property mapping by decomposing the global property of the entire structure into \emph{additive} atomic contributions
$${\bf \mathcal{P}} = \sum_i {\bf \mathcal{P}}_i.$$
Each atomic property $\mathcal{P}_i$ is trained \emph{locally}, with a computational cost independent of the size of the entire structure. This makes the mapping $ \mathcal{S} \mapsto {\bf \mathcal{P}}$ \emph{scalable}, allowing application to very large systems because the computational cost scales linearly with the system size~\cite{Goedecker99}.

In summary, the structure-property mapping is performed in two steps:

(i) Cropping $N$ atom-centered spherical fragments of radius $r_\text{cut}$ from the original structure
\begin{equation}\label{eq:crop}
\mathcal S \xrightarrow{\text{crop}} \left\{ \mathcal S_i \right\}, \quad \text{ such that } \bigcup_i \mathcal{S}_i = \mathcal{S}
\end{equation}

(ii) Learning the atomic properties locally
\begin{equation}
\label{eq:Si2Pi}
\left\{ \mathcal {S}_i \right\} \xrightarrow{\text{learn}} \left\{ \mathcal {P}_i  \right\}, \quad \text{ such that } \sum_i \mathcal {P}_i = \mathcal {P}.
\end{equation}

A fragment
$$ \mathcal S_i \equiv \left\{(Z_j, {\bf R}_j) \Big| \|{\bf R}_j-{\bf R}_i\| \leq r_\text{cut}  \right\} $$
should be large enough to thoroughly contain the geometric information of the environment of atom $i$, enabling distinction from others. However, smaller fragments are preferable because both the transferability and scalability of the model critically depend on the \emph{locally machine-learnable} nature of $ \mathcal S_i \mapsto  \mathcal P_i$. The optimal value of $r_\text{cut}$ varies for different materials and depends on the locality character of the target property.

These fragments may overlap; that is, an atom can belong to multiple fragments. However, if $\|{\bf R}_j - {\bf R}_i\| > 2r_\text{cut}$, then $\mathcal{S}_i \cap \mathcal{S}_j = \emptyset$. An appropriate mathematical representation (called a descriptor) of the local environments $ \mathcal S_i$ for predicting an atomic property $\mathcal P_i$ must be invariant to the spatial orientation of the whole structure. Additionally, it should be independent of the inclusion or removal of atoms near the fragment boundary~\cite{Li16}.
Various methods satisfy these invariances, including atom-centered symmetry functions~\cite{Behler07},
many-body tensor representations~\cite{Huo},
spectrum of the overlap matrix~\cite{Sadeghi13,Li16},
smooth overlap of atomic positions (SOAP)~\cite{soap},
atom-density representation~\cite{Willatt19}
and many others~\cite{Musil21}.
These methods successfully encapsulate the geometric and chemical information required for learning the property $\mathcal P_i$ around atom $i$.

\subsection{Machine Learning}\label{sec:methods}
To implement Eq.~(\ref{eq:Si2Pi}), we design an ML scheme that takes atomic fingerprints as input and predicts atomic LDOS as output.
Our implementation
utilizes  the \textit{scikit-learn} Python module~\cite{scikit-learn}
and is based on and partly similar to the Gaussian process framework 
developed by Ben Mahmoud~\emph{et al.}~\cite{benmahmoud}  which minimizes    
\begin{equation}  \label{eq:l2Di_split}   
\sum_\mathcal{S}   
\Big(\mathcal{D}^\text{ref}(\varepsilon) -  \sum_{i\in {\mathcal S}} \mathcal{D}_i(\varepsilon)\Big)^2,
\end{equation}  
where the inner sum runs over the atoms in structure $\mathcal{S}$
while the outer sum runs over the structures in the training set. 
This \emph{structural learning} strategy utilizes structural DOS, $\mathcal{D}^\text{ref}$, as the reference.  
However, to take advantage of the availability of local DOS, $\mathcal{D}_i^\text{ref}$, we instead minimize  
\begin{equation}  \label{eq:l2Di}   
{\ell}^2(\varepsilon) = \sum_\mathcal{S}   
\sum_{i\in {\mathcal S}}   
\Big( \mathcal{D}_i^\text{ref}(\varepsilon)-\mathcal{D}_i(\varepsilon)\Big)^2.
\end{equation}  
This \emph{atomic learning} strategy will be shown to improve the accuracy of predicting DOS-derived physical properties
if atom-projected DOS serves as the reference. 

Albeit using different references, the target of the regression in both approaches is the atomic LDOS as a function of energy
\begin{equation}  
\label{eq:DiML}   
\mathcal{D}_i(\varepsilon) =  \sum_{j} \kappa({\mathcal S}_i,{\mathcal S}_j) \, x_j(\varepsilon).  
\end{equation}  
This expression can be interpreted as an expansion of the target in terms of $m$ basis functions $x_j(\varepsilon): \mathbb{R} \to {\mathbb{R}}$ with coefficients $\kappa({\mathcal S}_i,{\mathcal S}_j)$. 
{(Alternatively, one may consider $x_j(\varepsilon)$ as the coefficients and $\kappa({\mathcal S}_i,{\mathcal S}_j)$ as the basis of expansion~\cite{benmahmoud}.)}  
 The machine learns 
\begin{equation}\label{eq:x}  
{\bf x}(\varepsilon) = \underset{{\bf x}:\, \mathbb{R} \to  \mathbb{R}^m}{\arg\min}   
\big\{{\ell}^2(\varepsilon) +   
\lambda^2  \langle {\bf x}(\varepsilon)| {\bf x}(\varepsilon) \rangle  
\big\}  
\end{equation}  
from the $m$ atomic environment samples of the training set
as a ridge regression problem 
on a grid of energy values $\varepsilon \in [\varepsilon_\text{min}, \varepsilon_\text{max}]$.
The hyperparameter $\lambda$ regularizes the summation and prevents overfitting to the training datapoints 
by suppressing the norm  
$$  \left \langle {\bf x}| {\bf x} \right \rangle =  
\sum_j\sum_{j'}  
x_j(\varepsilon)  x_{j'}(\varepsilon) \kappa({\mathcal S}_j,{\mathcal S}_{j'}).  $$  
The symmetric positive-definite kernel $\kappa({\mathcal S}_i,{\mathcal S}_j)$ quantifies the similarity between pairs of atomic environments, ${\mathcal S}_i$ and ${\mathcal S}_j$:
the more similar the environment of the objective atom $i$ to that of an atom $j$ in the training set, the higher the weight of contributing $x_j$ in determining the LDOS of atom $i$. 

Prediction error is reported as the average loss normalized 
to the standard deviation, $\mathcal{L}/\sigma$. 
Normalization enables comparison across different materials and atomic types. 
For a quantity $g(\varepsilon)$, for example, 
\begin{align}  \label{eq:normloss} 
\mathcal{L}^2 &= \left \langle \big(g-g^\text{ref}\big)^2 \right \rangle,
\nonumber \\
\sigma^2 & = \left \langle \big(g^\text{ref}-\langle g^\text{ref}\rangle\big)^2 \right \rangle
,\end{align}
where $\langle g\rangle \equiv (N \int d\varepsilon)^{-1} \sum_{k=1}^N \int g_k(\varepsilon) d\varepsilon $
takes the average over the energy range and $N$ samples.
For structural property $\mathcal P$, on the other hand,  $\langle \, \mathcal{P}\,  \rangle$
denotes an ensemble average over the structures in the set.

\subsection{Datasets and Training}
In this study  we use  four sets of structures previously generated  with DFT calculations.
As listed in Table.~\ref{tab:sets},
the sets contain pure silicon and carbon structures as well as Si-C and Sn-S-Se compound structures.
In addition to chemical diversity,
the sets also differ by the XC functional type used to generate their reference DOS. 
For the pure silicon structures of Set A and Set B,
the reference DOS is determined by broadening the single-particle energy eigenvalues 
calculated with the PBE~\cite{PBE} parametrization 
within the generalized-gradient approximation (GGA) of the XC functional.~\cite{benmahmoud}
The DOS of the SiC structures in Set C is  calculated by the HSE06~\cite{HSE06} hybrid functional.~\cite{NOMAD,*NOMAD-dataset}
It is known that,
by incorporating a fraction of the non-local exact exchange energy,
   such hybrid functionals more accurately predict the electronic structure (and thus the DOS) of materials.
Finally, for the Sn-S-Se solid solution structures of Set D, the reference DOS was generated with 
the strongly constrained and appropriately normed (SCAN) meta-GGA  functional~\cite{SCAN}
which results in almost the same shape of DOS of HSE06 computations
  with a uniform shift of the unoccupied bands.~\cite{solidsolution}

Our machine learning procedure consists of the three standard steps, 
    i.e. training, validation and testing.
As training, we minimize the prediction error (loss) with respect to the model parameters 
for a given training set of input-output pairs.
Then, validation is performed on a separate (unseen) set of input-output pairs 
and the model hyperparameters are tuned to optimize its accuracy and  performance.
Instead of a single step training-validation,
we employ a five-fold cross-validation strategy, 
   in which the training set of input-output pairs is divided into five groups. 
In each iteration, four groups are used for training while one group is reserved for validation. 
This process is repeated five times, ensuring that each group serves as the validation set once.
The trained model is finally tested on a completely unseen test dataset. 
This step assesses how well the model generalizes to new data
and how it will perform in future applications.
For example, for the pure silicon structures, 
the machine is first trained and 5-fold cross-validated on Set A
and then is tested on Set B. 
This helps us to verify the scalability of the model by
  learning from small structures (Set A) and applying to large structures (Set B).
  The aim of the next test, which is performed on Set C consisting of both pure Si and C as well as compound Si-C structures, is to verify the transferability of the model.
 As reported in Table.~\ref{tab:sets},
Set C is randomly split into two parts with a ratio of 4:1.
The larger portion is used for training and 5-fold cross validation
while the smaller one is left for testing the trained model.
A similar splitting strategy is utilized in the experiment on Set D, 
  which aims to assess the generalizability of the LDOS model to chemically complex structures.
Note that Set D encompasses a diverse range of solid solution structures of the ternary compounds
Sn(S$_{1–x}$Se$_{x}$), Sn(S$_{1–x}$Se$_{x}$)$_2$ and Sn(S$_{1–x}$Se$_{x}$)$_3$,
generated previously to study their thermoelectric and optoelectronic applications.~\cite{solidsolution}

\begin{table*}[]
\caption{\label{tab:sets} Description of the datasets used for machine learning
	with total and atom-projected DOS from previous DFT calculations.}
\begin{tabular} {p{1.2cm}  l  c c } 
\hline\hline 
 & & \multicolumn{2}{c}{No. env.}  \\ \cline{3-4}  
Name & Description     & \;\; train \; & \; test \;  \\ \hline
Set A & 1,039 molten, amorphous, distorted diamond and $\beta$-tin Si structures~\cite{benmahmoud} &  29723 \\
Set B    & 25 large amorphous Si structures~\cite{benmahmoud} & & 5616   \\
Set C    & 15 Si and 17 C structures, 10 Si-C compound structures~\cite{NOMAD,*NOMAD-dataset} & 498 & 128 \\  
Set D    &  1,600  Sn$_n$(S$_{1–x}$Se$_{x}$)$_m$ solid solution structures~\cite{solidsolution} & 27000 & \, 5400 \\
\hline \hline 
\end{tabular}
\end{table*}

\section{Results and Discussion \label{sec:results}}
\subsection{Locality of LDOS}

\begin{figure}[t]
  \includegraphics[width=.9\columnwidth]{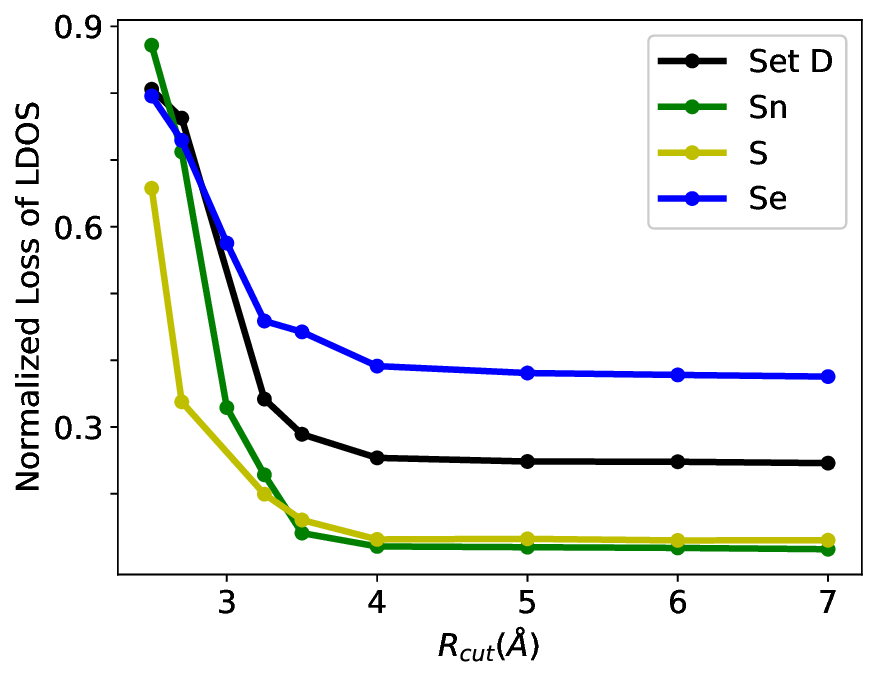}
  \caption{
  \label{fig:locality}
  Prediction error of LDOS as a function of the cropping radius for
 complex structures of Sn-S-Se solid solutions (Set D). 
 Reference LDOS is the projected DFT DOS.
 The normalized loss $\mathcal{L}/\sigma$ is averaged over the environments of each dataset.
    }
  \end{figure}

Unraveling the local learnability of atomic LDOS, a key assumption of this work, constitutes the first step of our investigation. 
Atomic LDOS is assumed to be locally learnable if the prediction error rapidly approaches zero 
when the radius of the cropped fragment around the corresponding atom increases; see Eq.~(\ref{eq:crop}). 
Our results indicate that as the fragment is enlarged, the prediction error decreases rapidly at first but then levels off. 
This behavior is observed across all tested atomic species
and is demonstrated in Figs.~\ref{fig:locality}
for selenium, sulfur and tin atoms in the solid solution structures of Set D.
We attribute the initial rapid reduction of prediction error below a cutoff radius of 4~\AA~  to the \emph{local machine learnability} of LDOS.
In other words, the LDOS assigned to an atom appears to adhere to the \emph{nearsightedness} principle of electronic matter,~\cite{Kohn2,Prodan05, *Prodan06}
and is practically learnable from the fragment that includes only a few neighboring atoms.

Beyond the locality regime, the prediction error becomes independent of fragment size and stabilizes at a finite value that varies by atomic species. 
For example, for silicon and carbon in Set C, the error saturates at approximately $\mathcal{L} \sim 0.5~\sigma$. 
A similar saturation value is observed for selenium in the ternary compound, as seen in Fig.~\ref{fig:locality}.
In contrast, the LDOS of sulfur and tin atoms is learned with greater accuracy, within $0.12~\sigma$.

Note that,
despite the absence of a true experimental reference DOS for training and testing,
the reference DOS for Set C and Set D is  generated 
using hybrid (HSE06) and meta-GGA (SCAN) XC functionals.
A potential non-local effect of such XC functionals could lead to  
a size-dependent reduction of prediction error  
 if the fragment extends over several interatomic distances.
 However, detecting this effect is hindered by 
the saturated prediction error observed beyond the locality regime.
Further systematic tests could clarify whether  
the observed local learnability of DOS is totally an inherent property of the electronic structure of materials 
or if artificial factors also play a role. 
Nonetheless, our findings suggest that, in practice, the LDOS can be effectively learned locally, 
with accuracy constrained by the saturated error.
It should also be noted that 
due to the high computational cost of DFT computations using hybrid functionals, 
 local or semilocal XC functionals, such as those used for Set A and Set B,
 are commonly employed in electronic structure calculations. 
Furthermore,  as mentioned in Sec.~\ref{sec:LDOS}, 
 unlike $\mathcal{D}(\varepsilon ,{\bf r})$, 
 atomic $\mathcal{D}_i(\varepsilon)$ 
 is not a directly measurable physical quantity.
 Instead, it is derived from an arbitrary projection of the DFT wavefunction onto atomic orbitals or within atomic regions (e.g., Wigner-Seitz cells).
 We will not elaborate further on this  issue here. 
 In the following results, we set $r_\text{cut} = 6$~\AA.

  \subsection{DOS-Dependent Physical Properties}
The electronic DOS is not typically the ultimate quantity of interest; however, it is essential for calculating the structural properties of materials.
As illustrated in the following examples,
the accuracy of predicting physical properties from the learned atomic LDOS is fairly high. 
For convenience, the expressions that relate physical quantities to structural DOS are reviewed in Appendix~\ref{app:formula}.
Using those equations, we calculated the quantities for the largest dataset, set A, which contains 1039 silicon structures. 

\begin{figure*}[t]
    \includegraphics[width=.49\columnwidth]{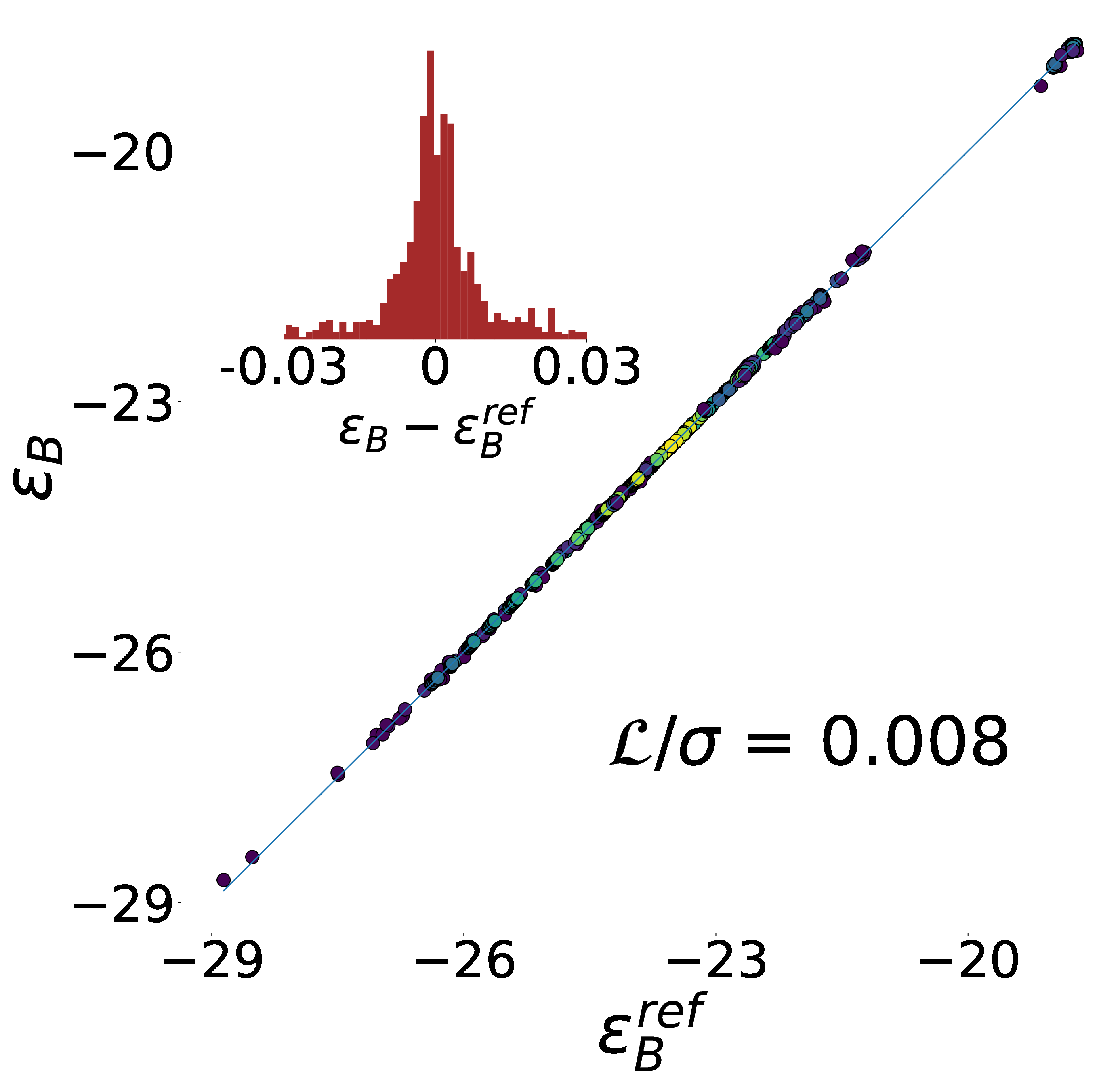}
    \includegraphics[width=.49\columnwidth]{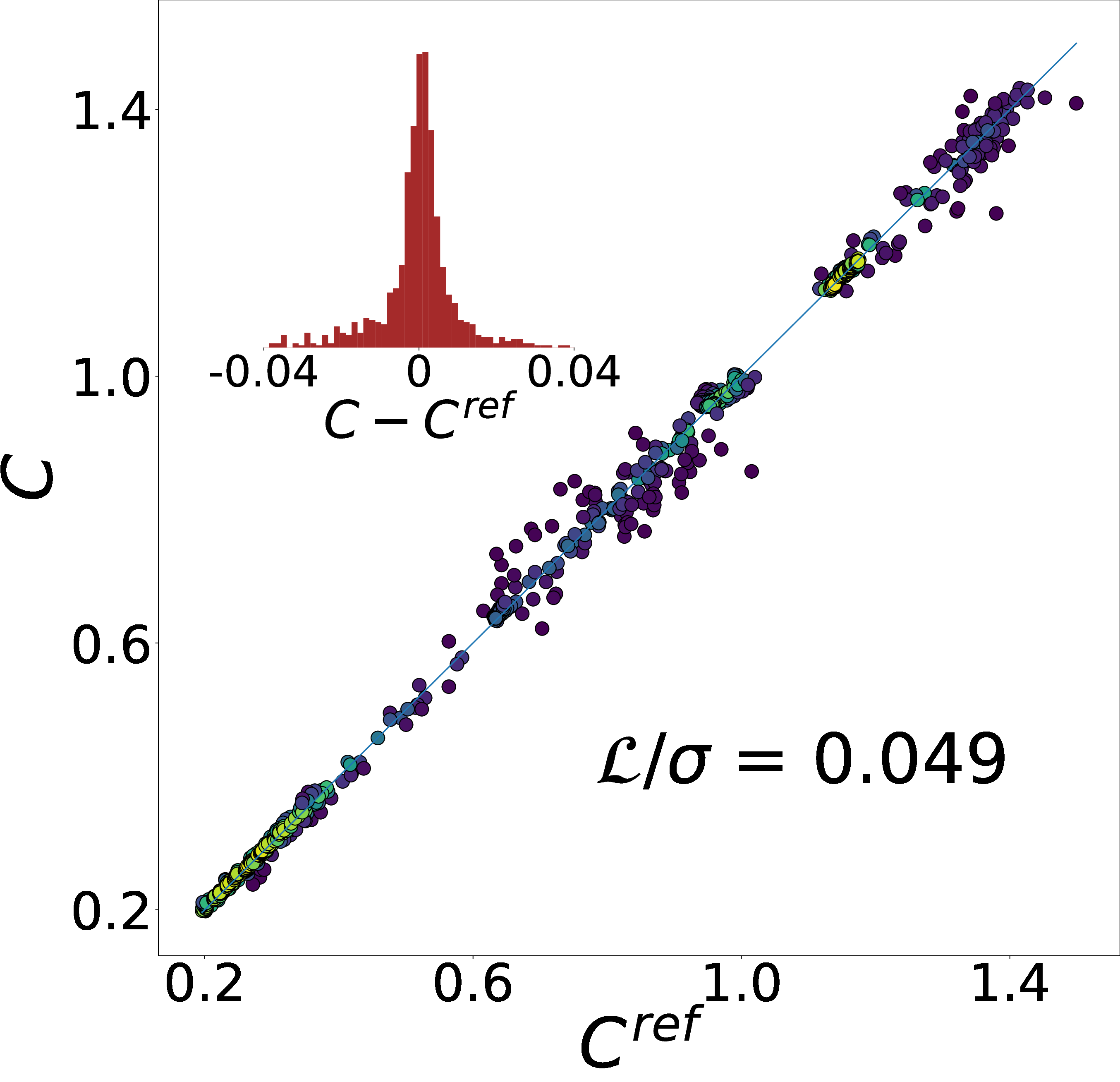}
    \includegraphics[width=.49\columnwidth]{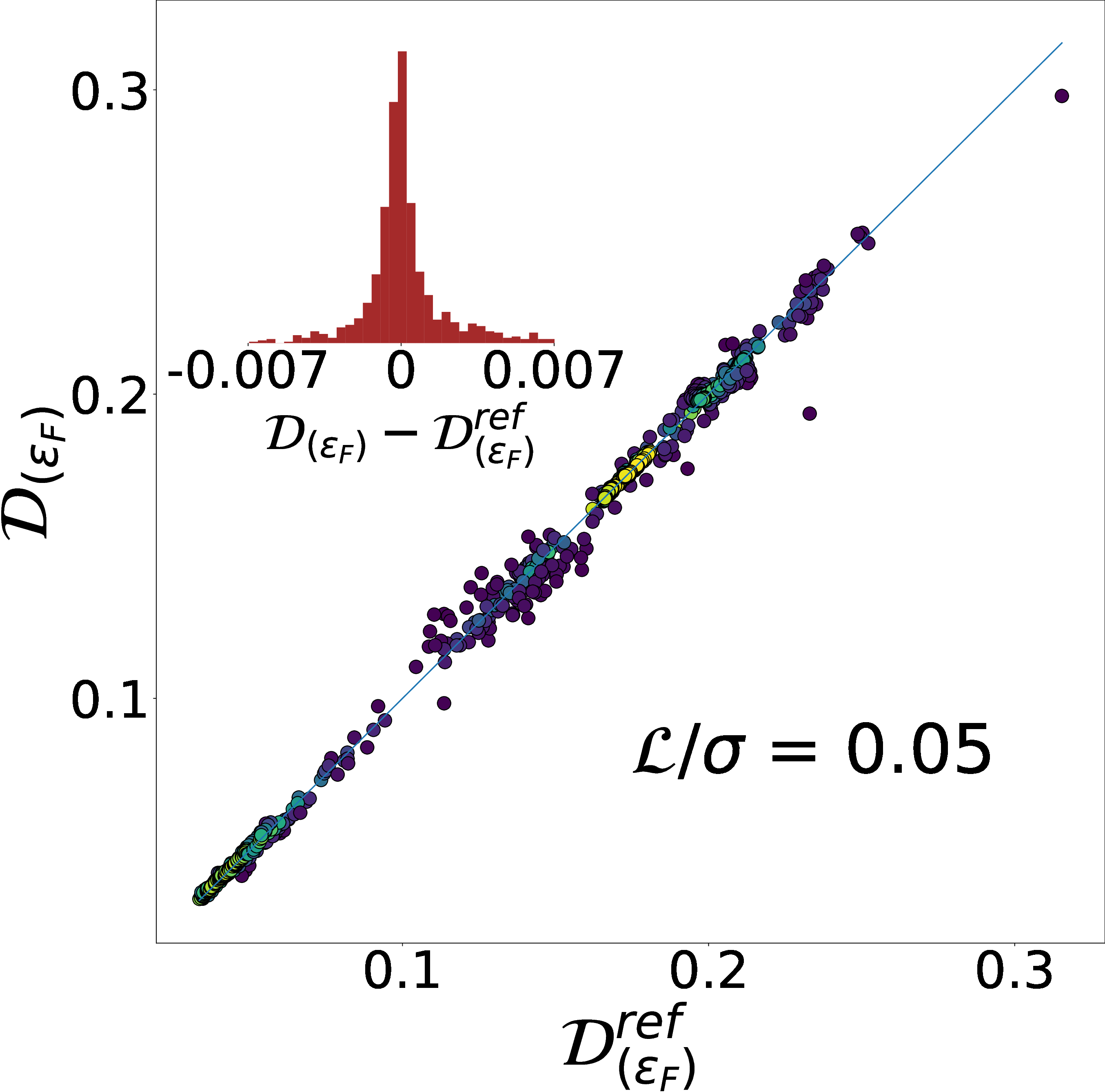}
    \includegraphics[width=.49\columnwidth]{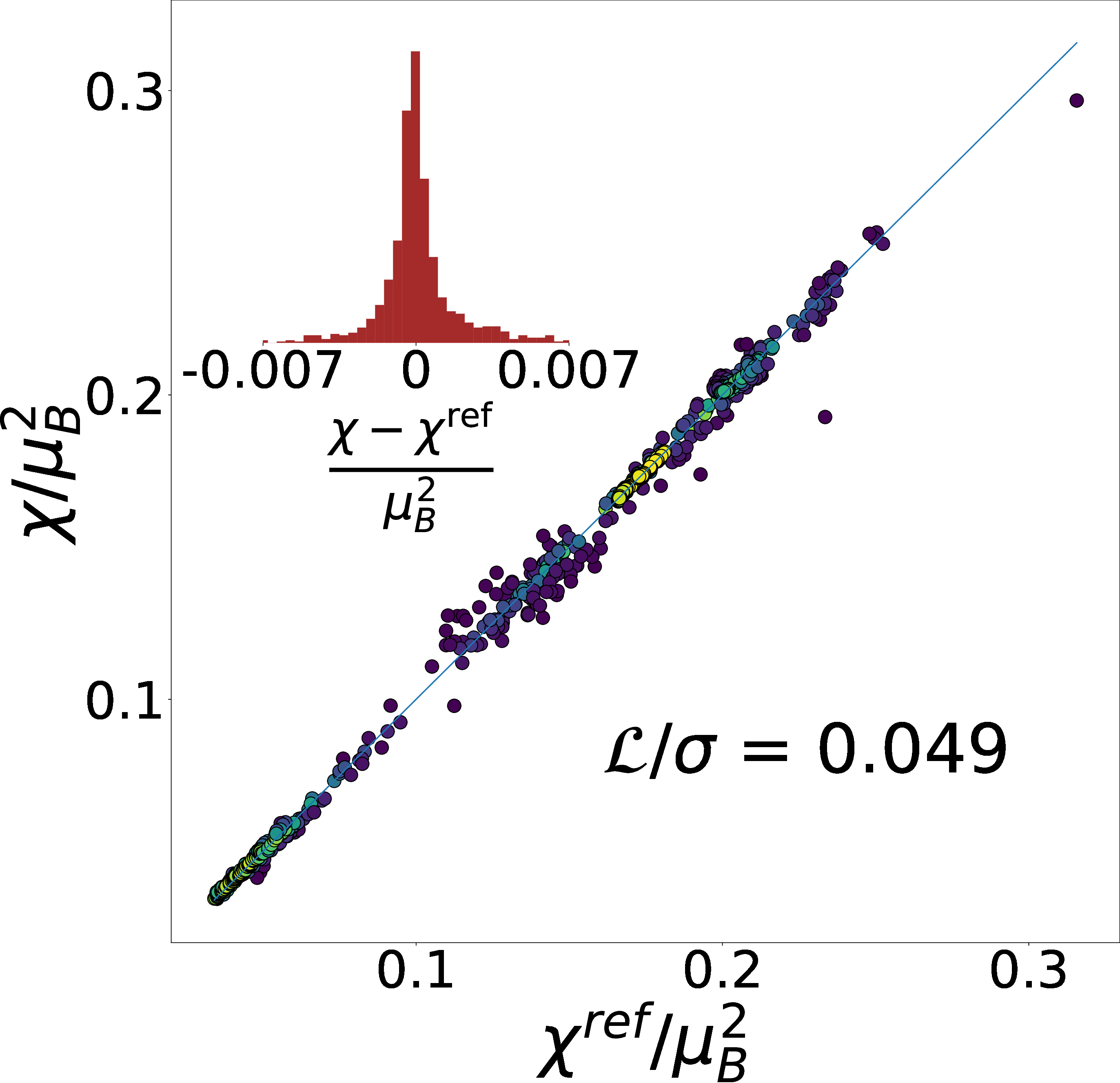}
    \caption{
    \label{fig:scatter}
	Parity plots of predicted physical quantities derived from the predicted DOS versus DFT reference values for 1039 silicon structures (Set A):
	band energy (in eV), heat capacity (in eV K$^{-1}$), DOS at the Fermi level (in eV$^{-1}$), and magnetic susceptibility. The diagonal line indicates perfect prediction. 
Inset: The corresponding histograms of prediction errors.
    }
\end{figure*}

The parity plots of the ML-LDOS-derived quantities 
against the reference values calculated from the DFT DOS are presented in Fig.~\ref{fig:scatter}
for band energy, structural DOS at the Fermi level, electronic heat capacity, and magnetic susceptibility.
Note that error is defined as the loss averaged over the structures and normalized to the standard deviation.
The error histogram, shown in the insets, demonstrates a small variance with no bias in all cases.
Overall, the performance of the predictions is fairly good
with the band energy showing much less scattering from the perfect diagonal line.
One may also expect a very small prediction error for  $\varepsilon_b$ and $ C(T) =  \partial \varepsilon_b /{\partial T} $.
However, it appears that the numerical integration of the trained LDOS introduces a considerable error.  
On the other hand, 
the high similarity between the parity plots of \(\mathcal{D}(\varepsilon_F)\) and \(\chi\) can be explained as follows. The Fermi-Dirac distribution function changes significantly only in the vicinity of the chemical potential, where its derivative 
exhibits a sharp, Dirac delta function-like peak. 
In particular at low temperatures, 
  $\partial f_\text{FD} /\partial\varepsilon \simeq  -\delta(\varepsilon-\varepsilon_F)$ 
  vanishes except close to the Fermi level $\varepsilon_F$ so that 
 the integral in Eq.~(\ref{eq:chi}) can be approximated as 
  $ \chi \approx \mu^2_B \mathcal{D}(\varepsilon_F)$. 

\subsection{Learning from Local versus Total DOS}

\begin{figure}[t]
    \includegraphics[width=0.51\columnwidth]{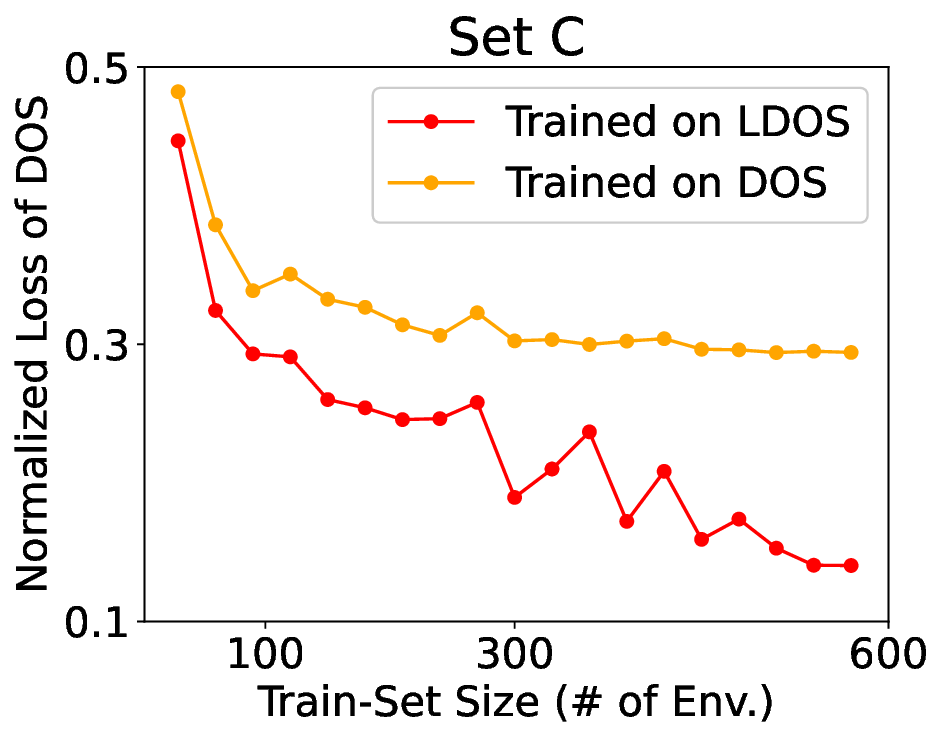}
    \includegraphics[width=0.47\columnwidth]{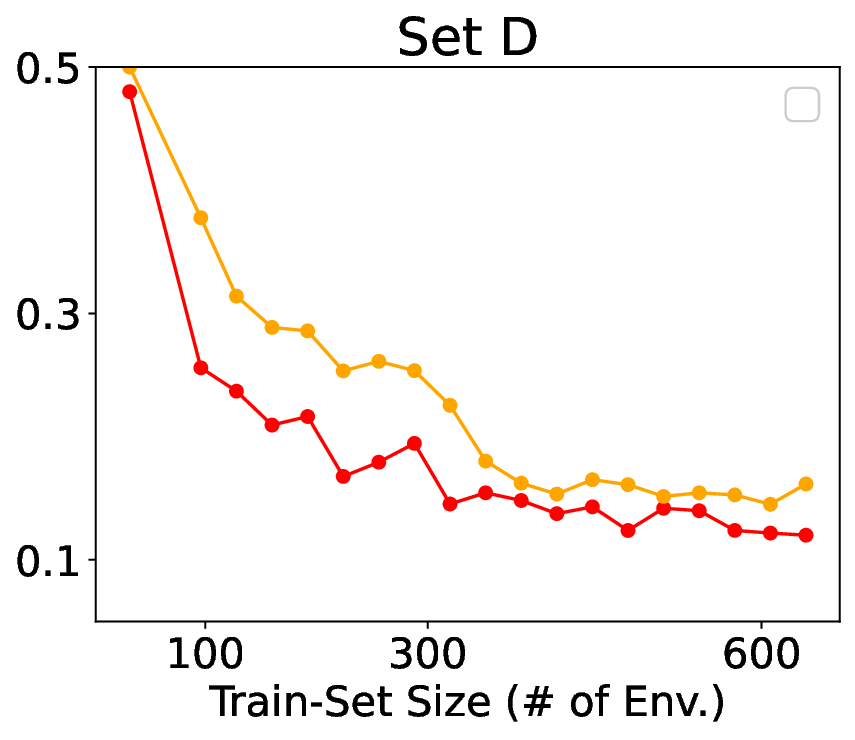}
    \caption{Learning curves for structural DOS, $\mathcal{D} =\sum_i \mathcal{D}_i$ 
	for the C-Si (Set C) and Sn-S-Se (Set D) structures.
       Learning prediction accuracy is compared for two strategies: training on the DFT LDOS from wavefunction-projection 
	or on the ML-split original DFT DOS. 
 \label{fig:lcurve}
   }
\end{figure}

As mentioned earlier, projecting the structural DOS onto atomic centers can be done in several ways, 
including projection onto atomic orbitals as Eq.~(\ref{eq:Di_a}) 
or splitting~\cite{benmahmoud} according to Eq.~(\ref{eq:l2Di_split}). 
In Fig.~\ref{fig:lcurve}, we compare the training on these two types of reference LDOS for the compounds (Set C and Set D).
The learning curves,
in particular for Set C containing Si and C environments, 
show a better performance when the atomic LDOS is trained on the physically meaningful projected DOS rather than on the total DOS.
In other words, learning from the atom-projected DOS instead of the total DOS leads to improved accuracy.
It is important to note that no significant additional computational effort is required to project the wavefunction onto atomic centers once the wavefunction is determined through a self-consistent procedure. Therefore, we propose replacing the LDOS-on-DOS training strategy introduced by Ben Mahmoud \emph{et al.}~\cite{benmahmoud} with an LDOS-to-LDOS training strategy. 

\begin{figure}
  \includegraphics[width=.75\columnwidth]{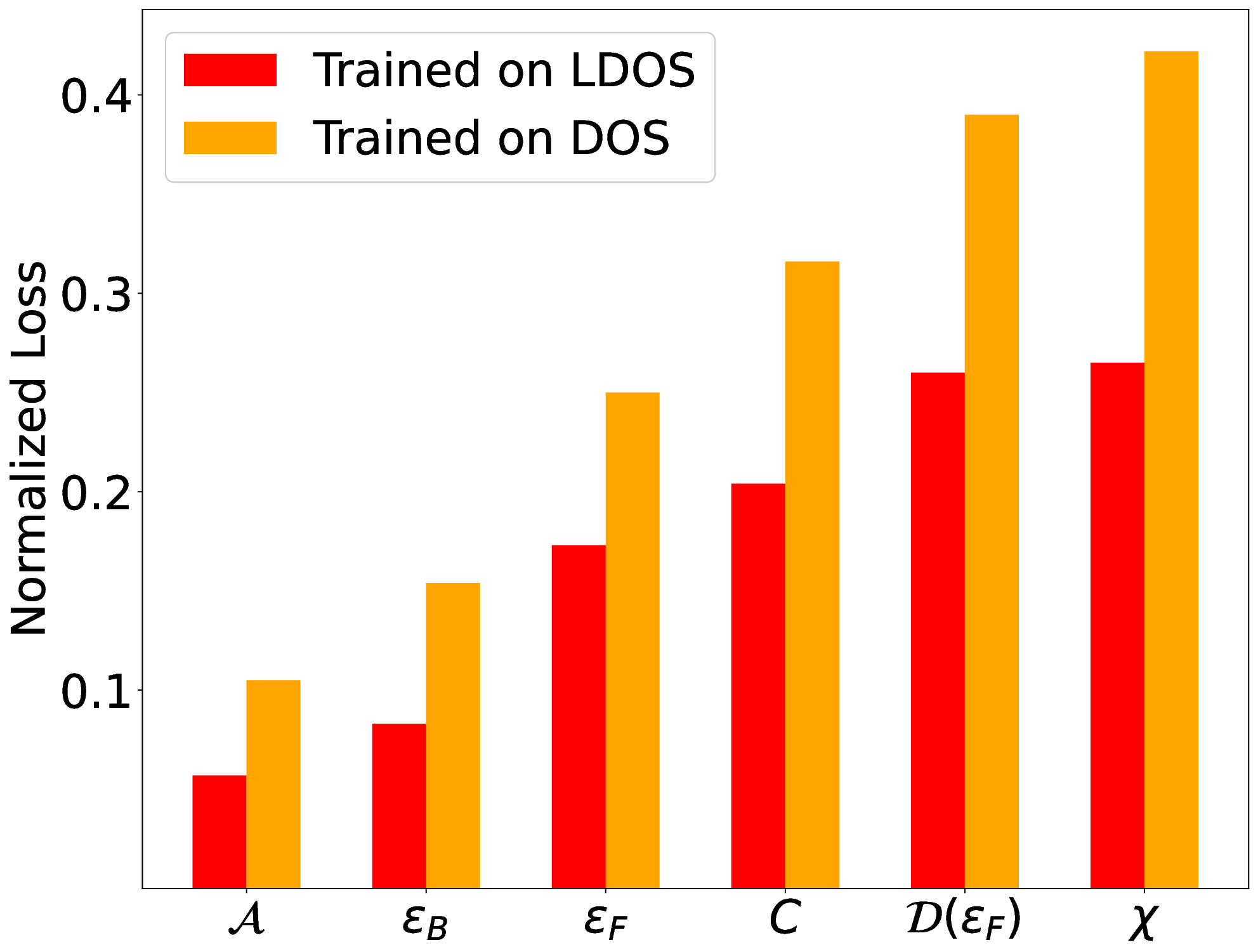}
  \caption{ \label{fig:prop}
Comparison of prediction error for several physical properties
   (band energy,   adsorption spectrum, electronic heat capacity, Fermi energy, DOS at the Fermi level, magnetic susceptibility)
	as obtained from the ML predicted DOS and LDOS.	
	 The normalized loss $\mathcal{L}/\sigma$  is calculated over the structures in Set C. 
           }
  \end{figure}

We compare the prediction accuracy of physical properties derived from our learned atomic LDOS 
with those derived from the learned structural DOS (i.e. the original approach by Ben Mahmoud~\emph{et al.}~\cite{benmahmoud})
in Fig.~\ref{fig:prop} for Set C.
The normalized loss is considerably  smaller in all cases when the property is derived from the ML-LDOS compared to the ML-DOS. 
The smallest error is observed for the adsorption spectrum, while the DOS value at the Fermi level and susceptibility exhibit the largest loss. 
Overall, the accuracy of calculating the physical quantities remains within 40\% of the standard deviation.

 \begin{figure}[t]
    \includegraphics[width=.50\columnwidth]{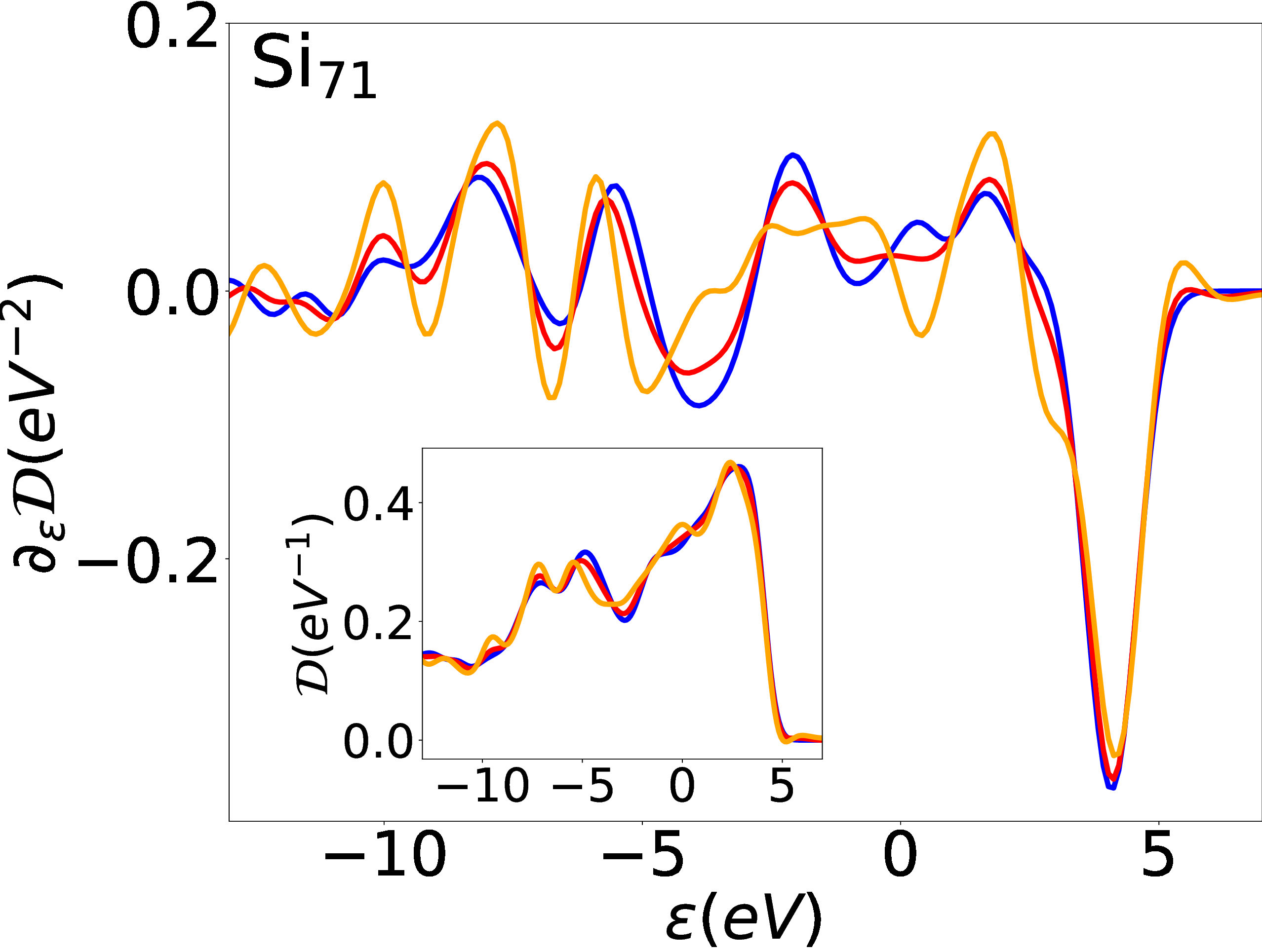} 
    \includegraphics[width=.47\columnwidth]{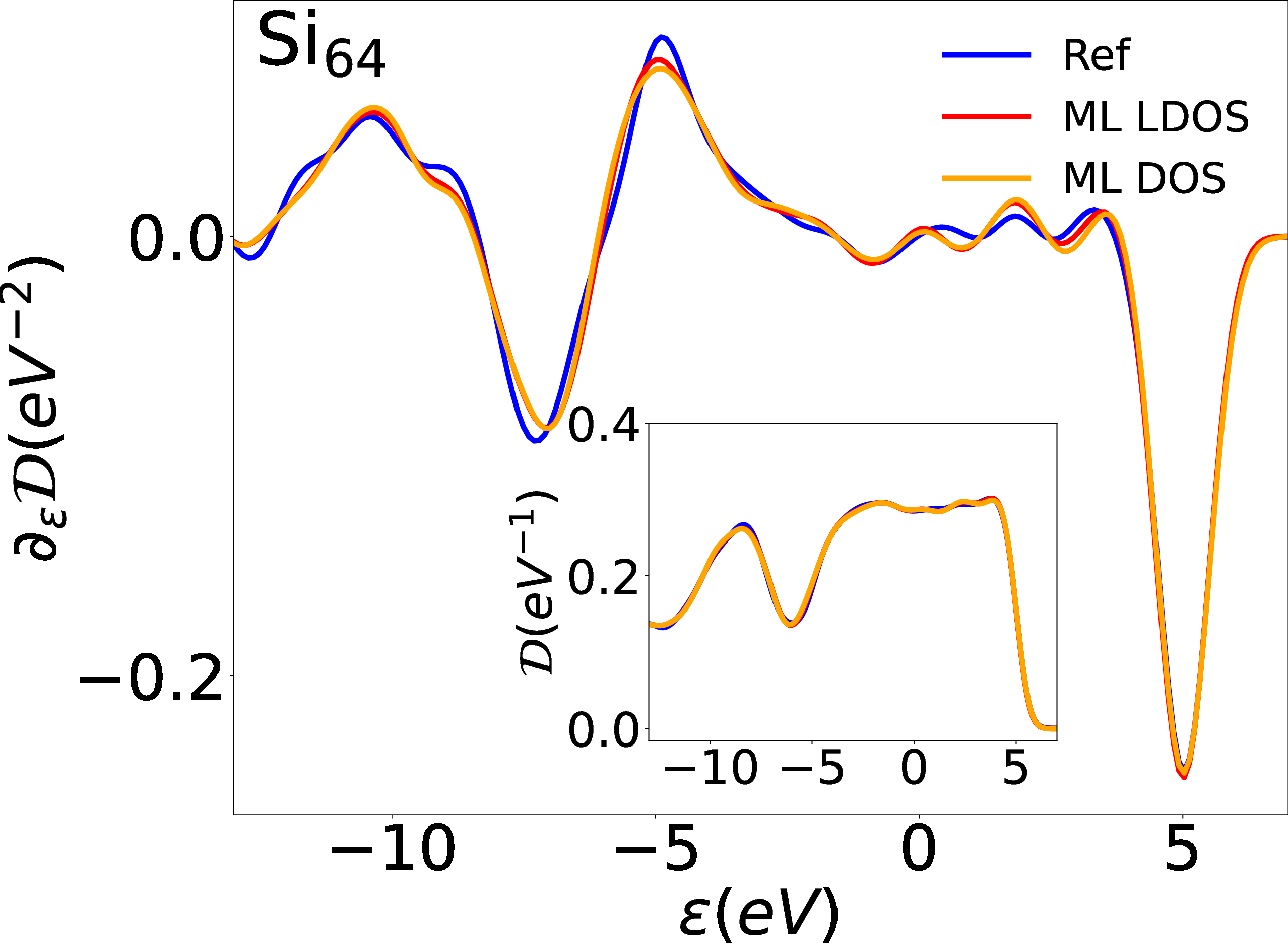} 
    \caption{ \label{fig:ddos}
        Derivative of structural DOS with respect to energy, $\partial\mathcal{D}/\partial \varepsilon$, as a function of energy 
	 for two representative structures,
	 when LDOS is trained on total DOS or on atom-projected contributions. 
  Inset: The corresponding DOS, $\mathcal{D}$. 
        For comparison, the DFT reference is shown by blue dashed curves.
  }
   \end{figure}

We have demonstrated that training on atomic LDOS enhances the accuracy of estimating derived physical quantities. 
These quantities can be significantly influenced by fluctuations in DOS, particularly near the Fermi level~\cite{nearfermi1,*nearfermi2}. 
The derivative of the DOS with respect to energy is illustrated in Fig.~\ref{fig:ddos} for two representative Si structures. 
It is evident that \(\partial \mathcal{D}/\partial \varepsilon\) can be predicted more accurately when the LDOS is trained compared to when the structural DOS is trained directly. 
Notably, the prediction error decreases when abrupt changes and fluctuations in the original (L)DOS are smoothed out prior to training. 
This is not surprising, as seen in Figs.~\ref{fig:dos_pure} and \ref{fig:dos_compound}, that 
the deviations from the reference curve are more pronounced at locations of sharp variations in the LDOS and DOS.
In fact, the ubiquitous incompleteness and finite temperature effects provide a natural smoothing of the sharp peaks in practice.  
To effectively achieve this smearing  and control the smoothness of the DOS, we can replace the Dirac delta function in Eq.~(\ref{eq:DOS0}) and subsequent equations with a Gaussian function 
$ \delta(\varepsilon-\varepsilon_n) \simeq 
(\sqrt{2\pi} k_B T)^{-1} \exp\left[-\left(\frac{\varepsilon - \varepsilon_n}{\sqrt{2} k_B T}\right)^2\right]$, 
which mimics an artificial temperature \(T\).
Our tests 
indicate that the normalized loss decreases by a factor of two if the artificial temperature increases from 0.1 to 0.5~eV.
Note that \(k_B T = 0.5\) eV in Fig.~\ref{fig:locality}.
We also found that the prediction error for LDOS is slightly smaller than that for the DOS,
with normalized losses of {$\mathcal{L}/\sigma$} = 0.07 and 0.09, respectively for $k_B T$=0.5~eV.

  \subsection{\label{sec:DOS} Generality of Local  Learning}


 \begin{figure*}
  \includegraphics[width=.20\textwidth]{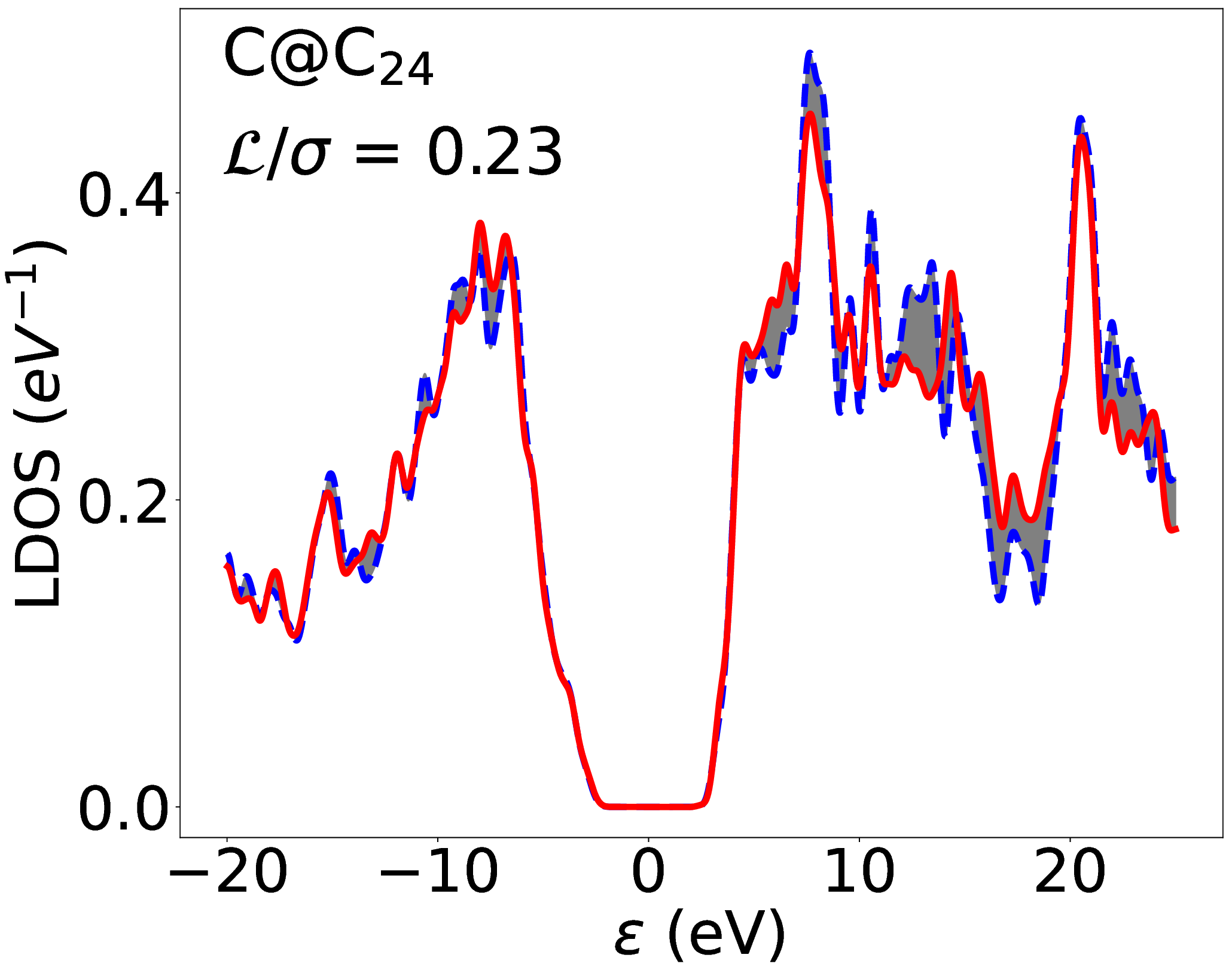} 
  \includegraphics[width=.19\textwidth]{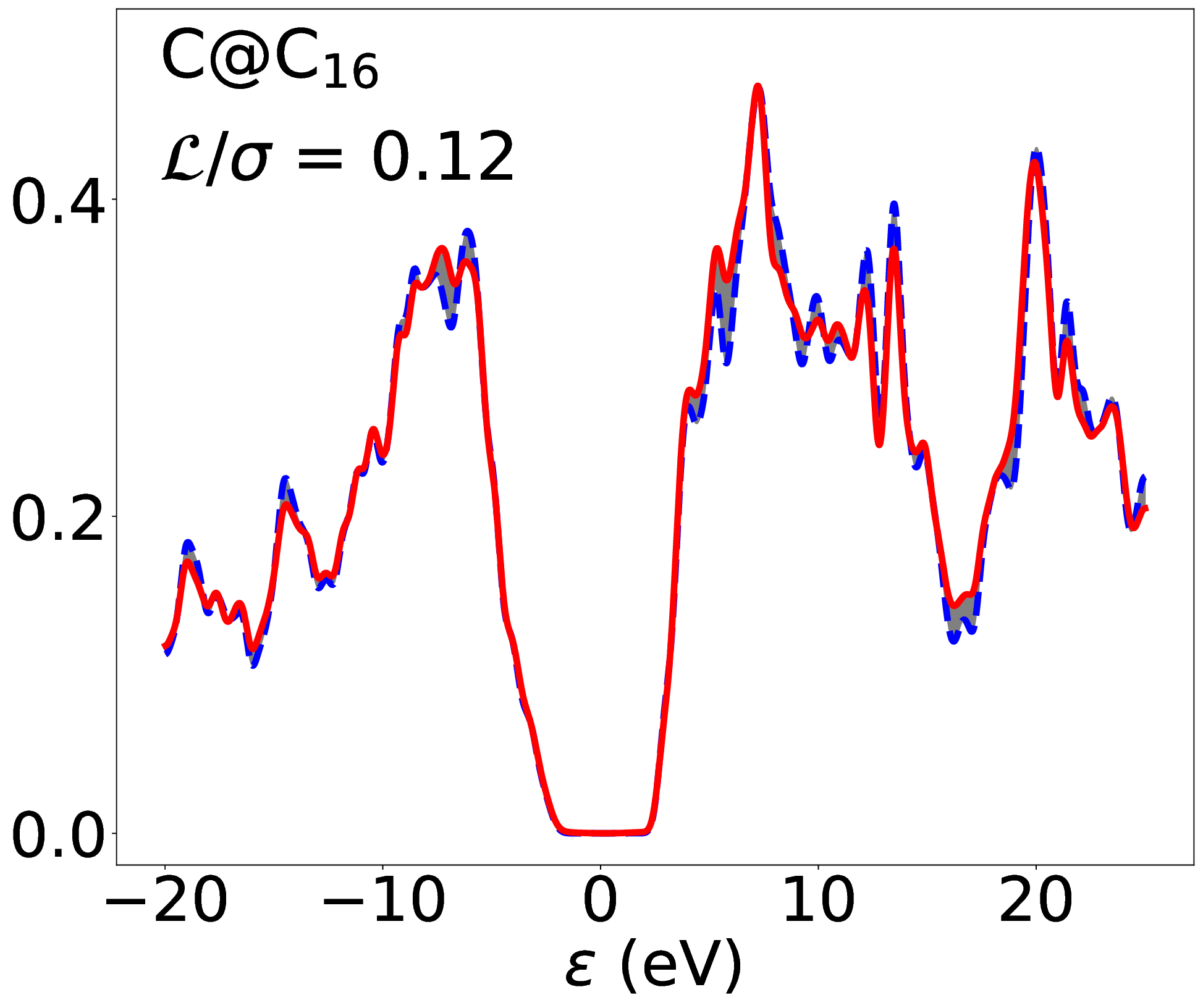} 
  \includegraphics[width=.19\textwidth]{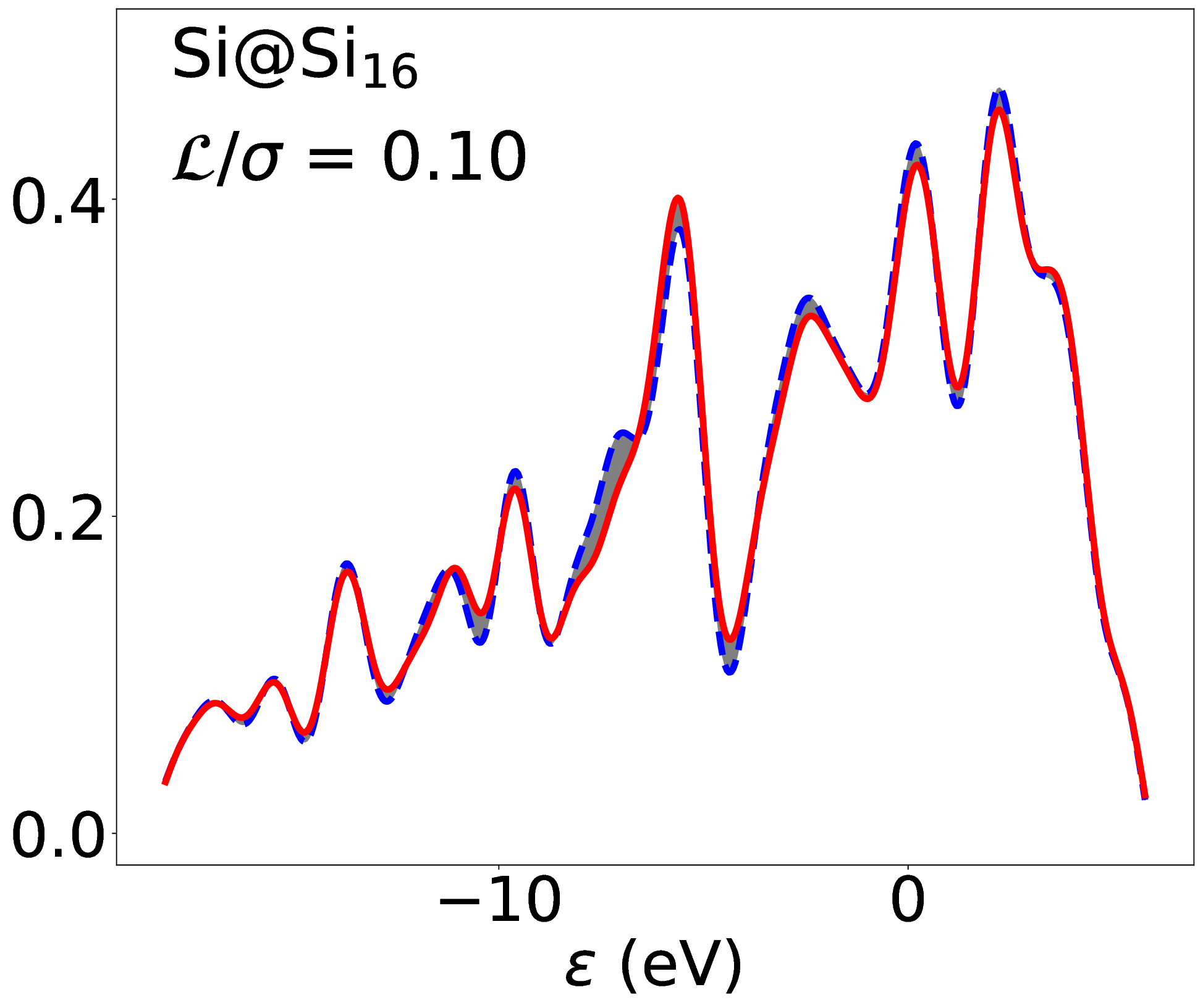} 
  \includegraphics[width=.19\textwidth]{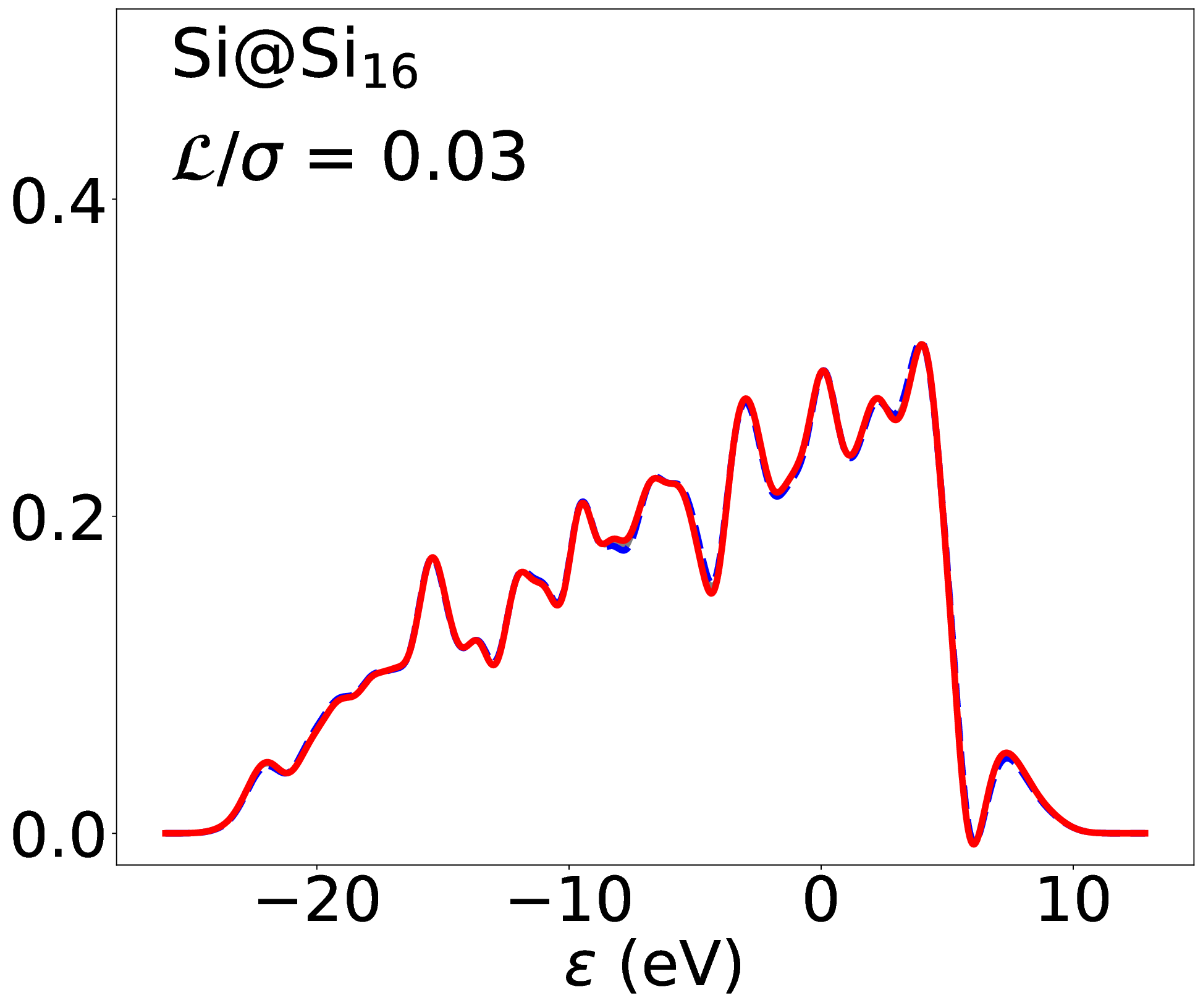} 
  \includegraphics[width=.19\textwidth]{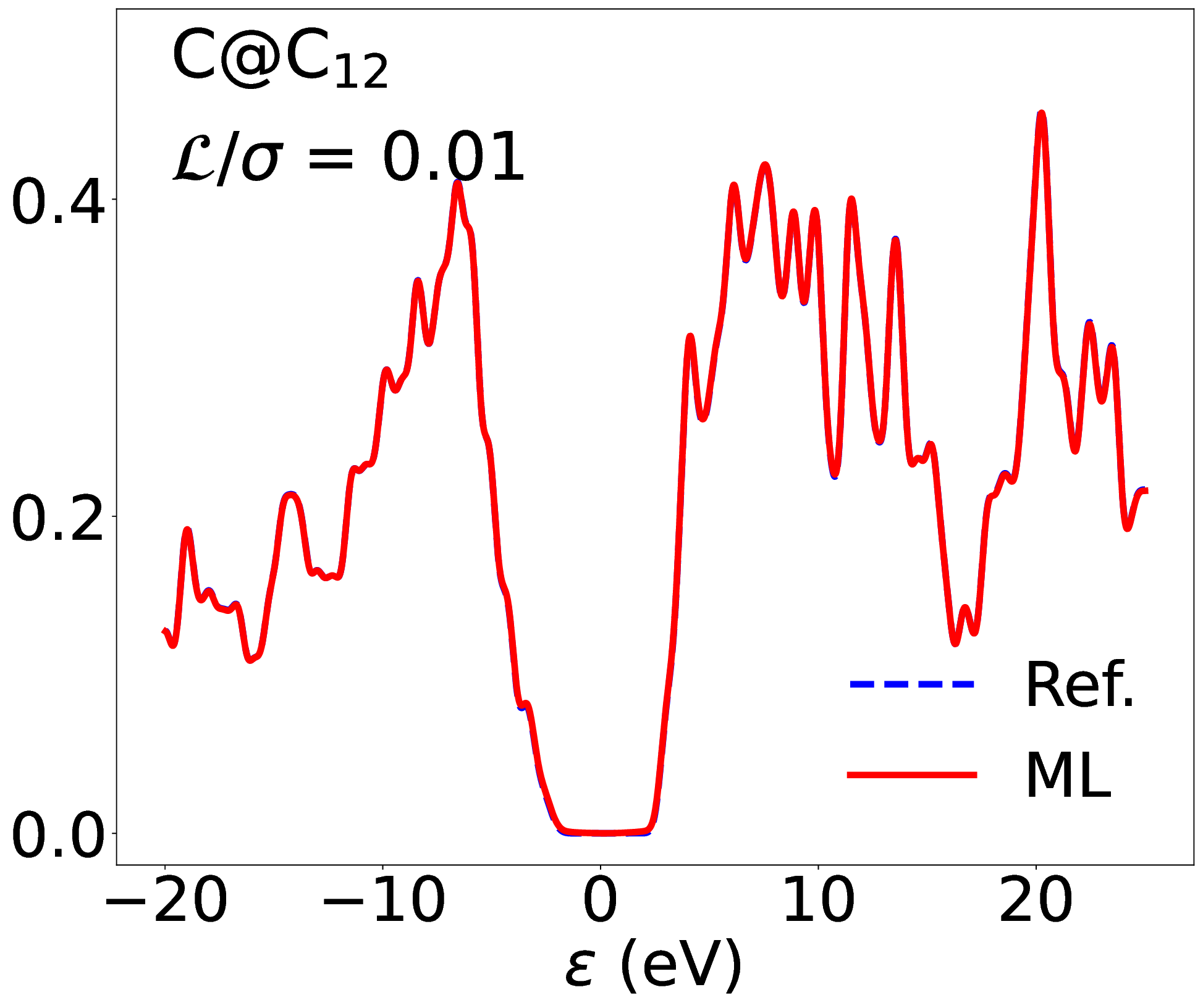} 
\\
  \includegraphics[width=.20\textwidth]{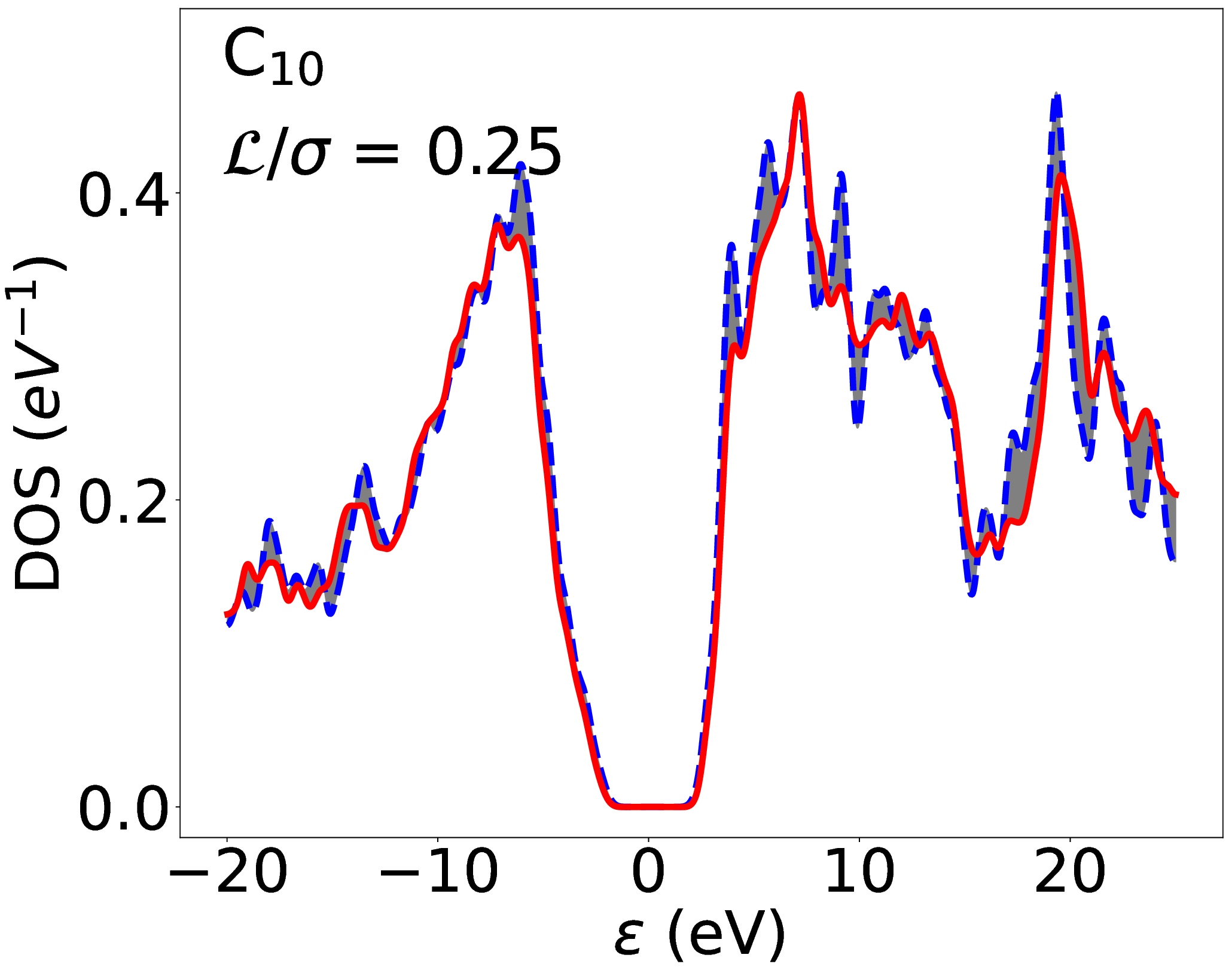} 
  \includegraphics[width=.19\textwidth]{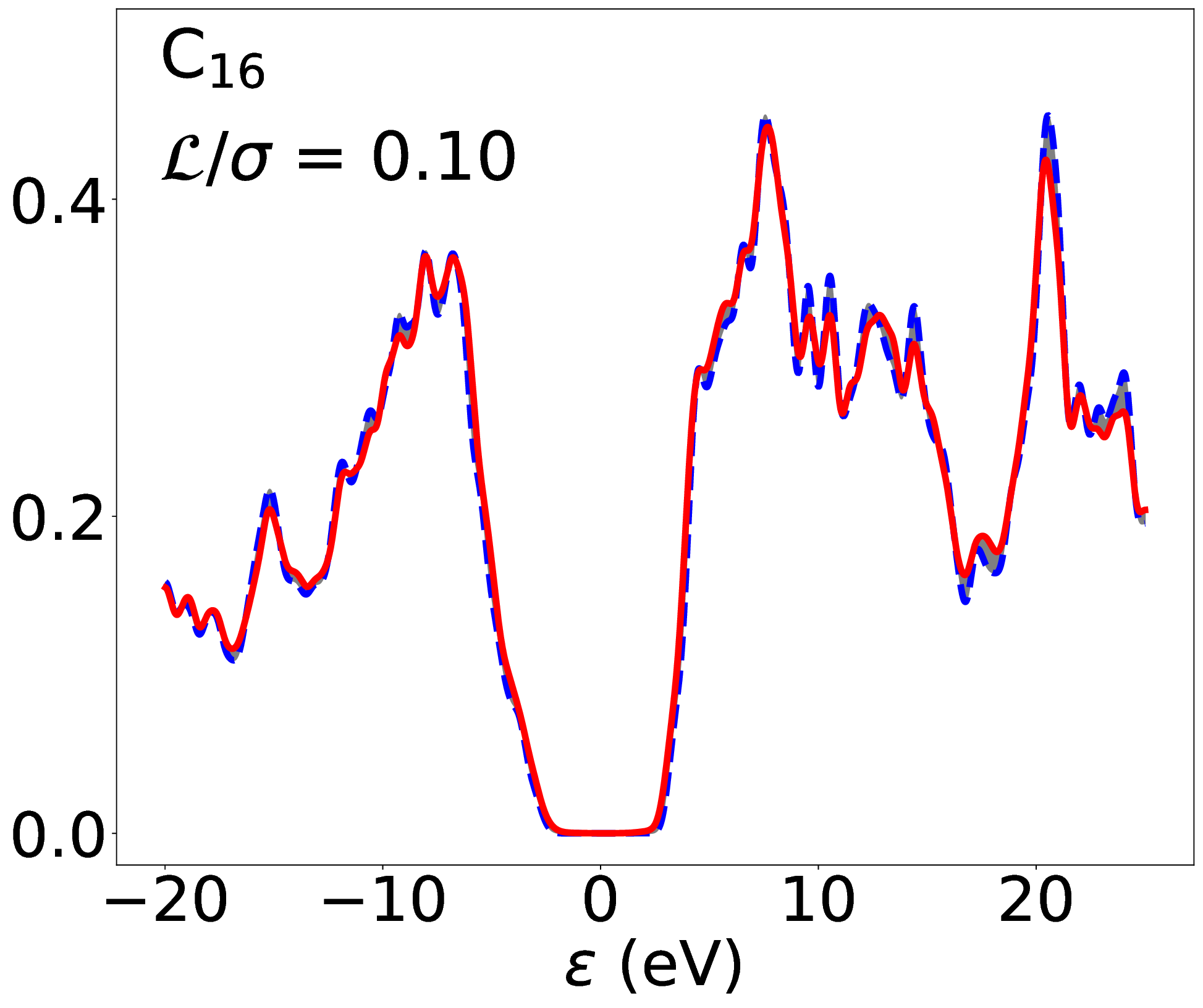} 
  \includegraphics[width=.19\textwidth]{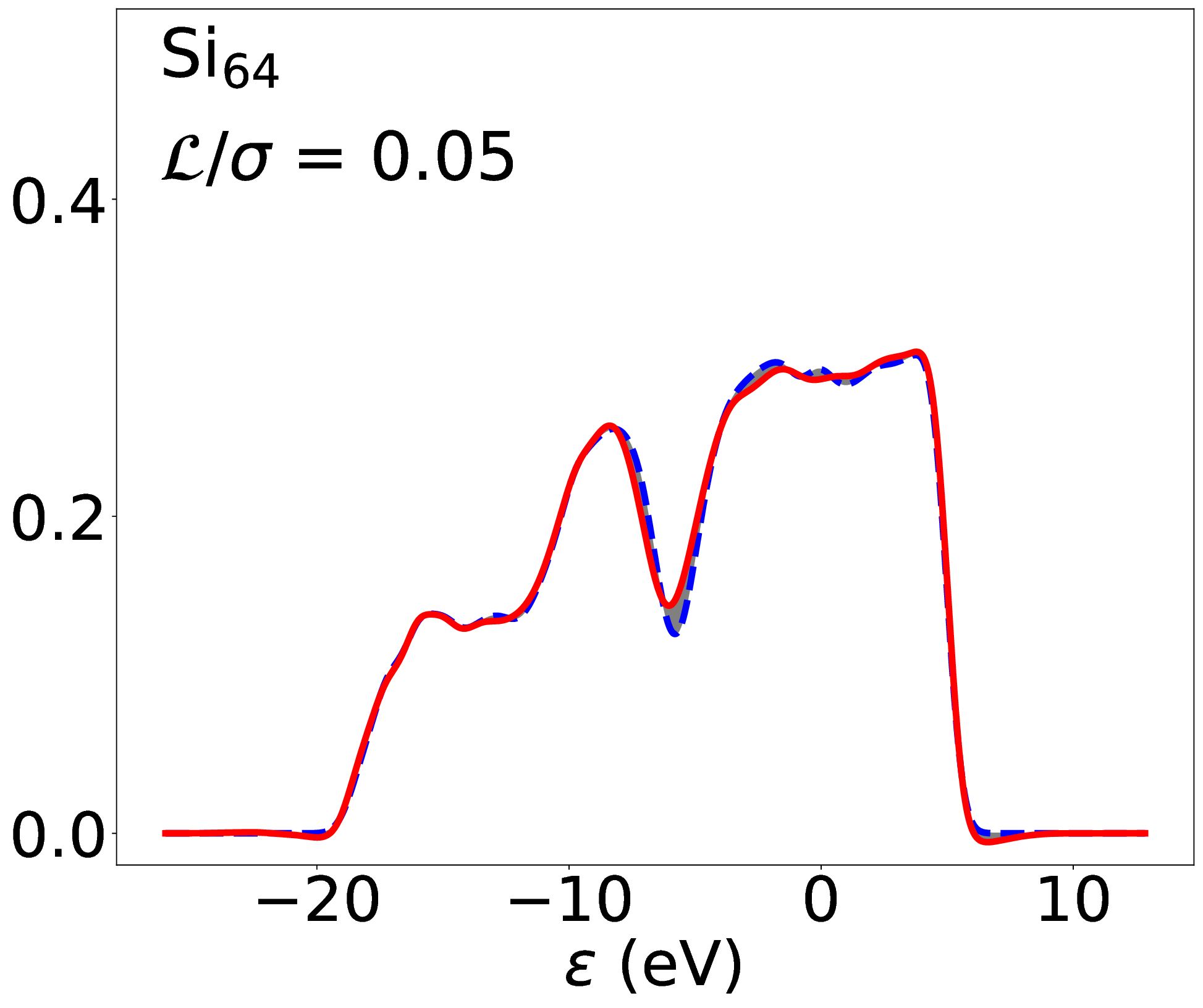} 
  \includegraphics[width=.19\textwidth]{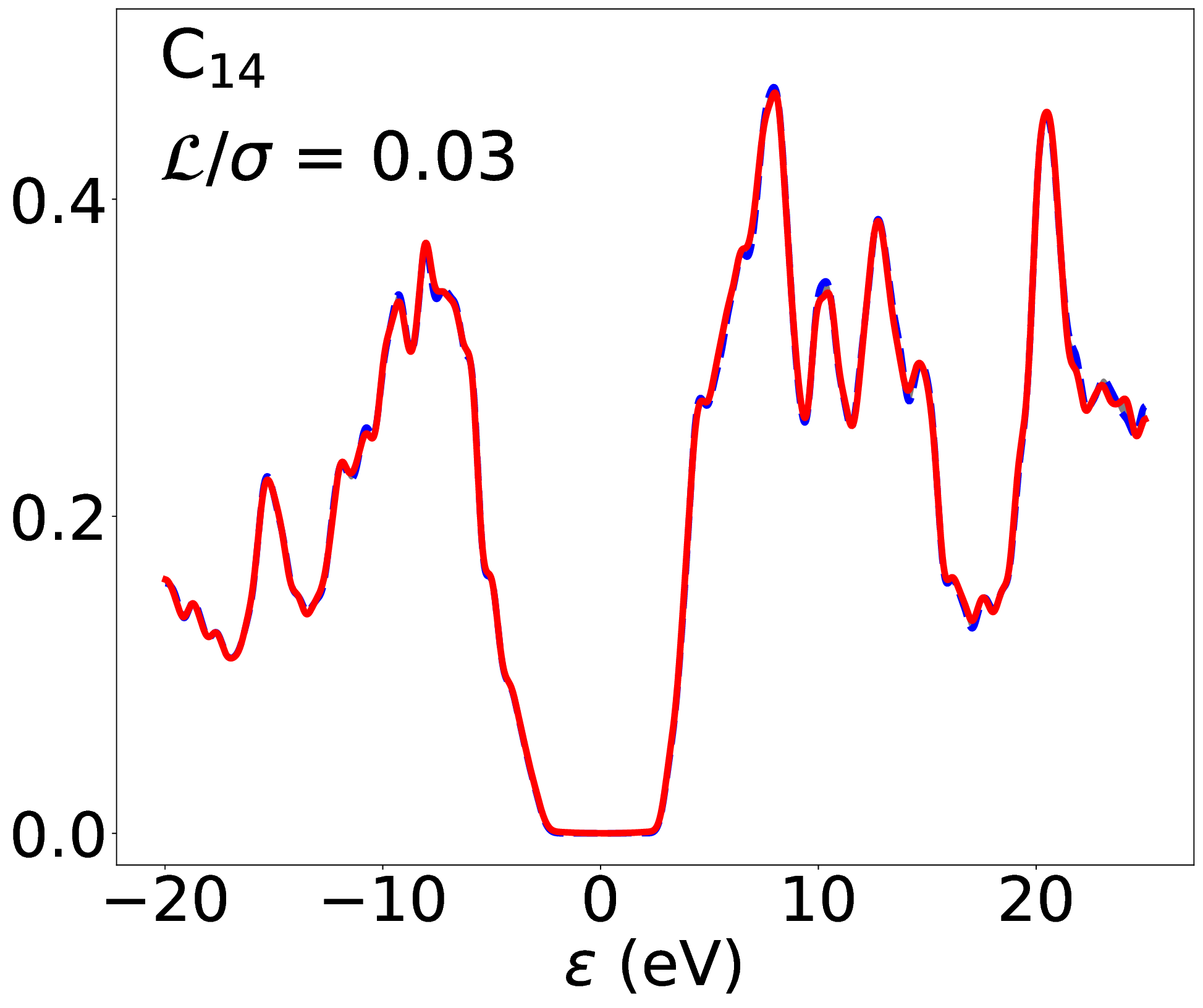} 
  \includegraphics[width=.19\textwidth]{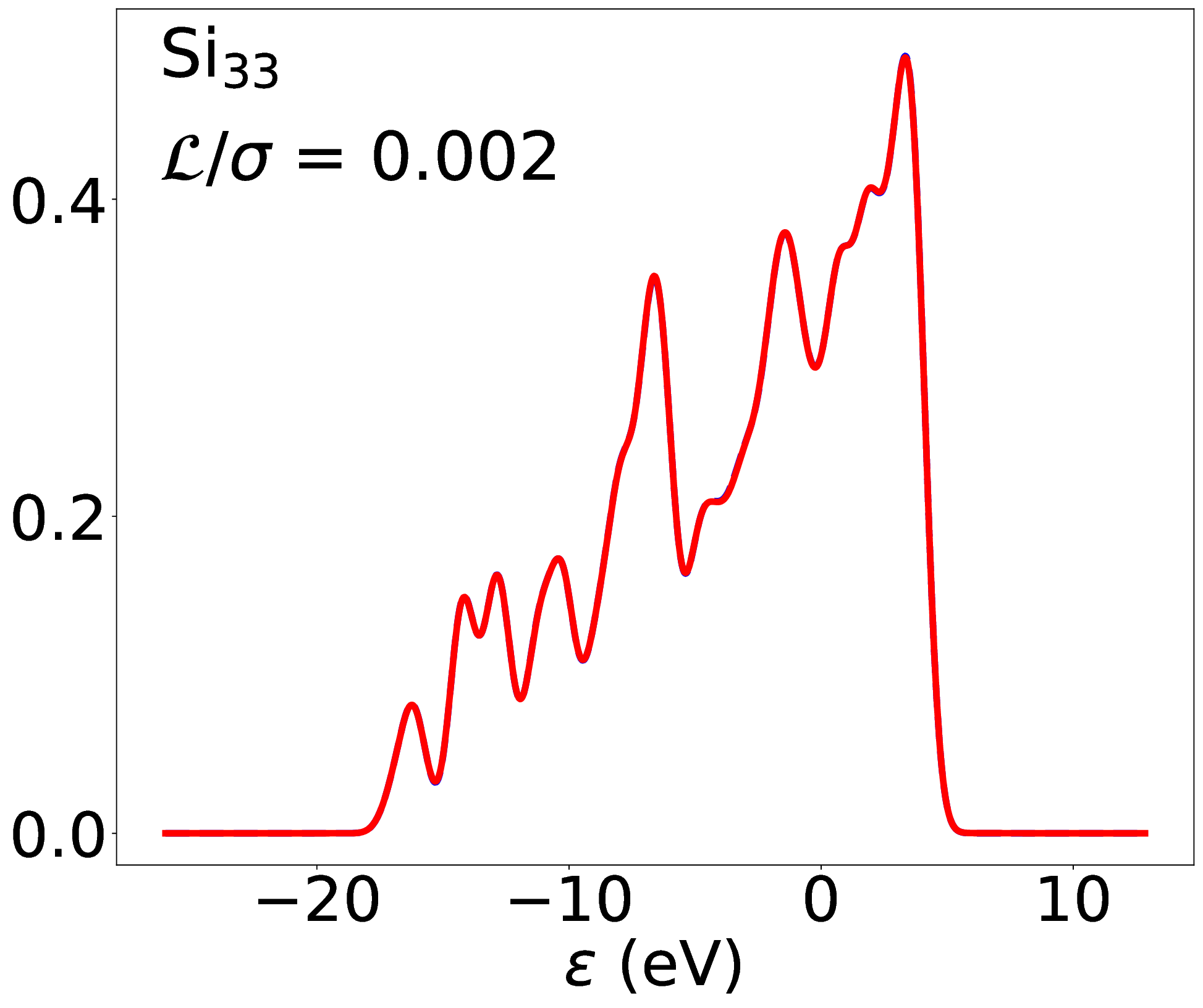} 
    \caption{\label{fig:dos_pure}
    Representative predicted atomic LDOS (top row) and structural DOS (bottom row) for pure silicon 
	 (from Set A) and pure carbon structures (from Set C).
	 Each row includes the worst, the median and the best prediction in the corresponding set,
	 and the panels are accordingly sorted 
	 from the maximum loss of LDOS of $\mathcal{L}$ = 0.23~$\sigma$ to the minimum one of 0.01~$\sigma$  (top row)
	 and from $\mathcal{L}$ = 0.25 to 0.002~$\sigma$ for DOS (bottom row).
        For comparison, the DFT reference is shown by blue dashed curves.
  }
  \end{figure*}

 \emph{Pure Structures.}
 The predicted LDOS for representative silicon and carbon atoms in pure structures (Set A and Set C) 
 is shown in 
 the top row of Fig.~\ref{fig:dos_pure}.
 For comparison, the predicted structural DOS for representative structures is also shown on the bottom row. 
The representatives are selected according to the prediction error and include the worst  
and the best predictions, shown on the rightmost and leftmost columns, respectively.
As before, the prediction error  is the normalized loss $\mathcal{L}/\sigma$; see Eq.~(\ref{eq:normloss}).
The  performance is remarkable for both LDOS and DOS predictions
 and a very good agreement with the reference curves is observed:
 the loss remains within 25\% of the standard deviation in all cases. 
The largest error in LDOS prediction accounts to $\mathcal{L}\sim 0.23~\sigma$ and corresponds to a carbon atom in a 24-atom pure structure.
The worst prediction of DOS, on the other hand, corresponds to a 10-carbon structure with a loss of $\sim 0.25~\sigma$.
For a 64-atom silicon cell (thermally randomized from a diamond crystal)  
a highly accurate DOS prediction is observed 
while an almost perfect prediction is gained for a Si$_{33}$ cluster structure.   
Recall that the structural DOS is calculated as the sum of the predicted atomic LDOS's.
Therefore,
 Fig.~\ref{fig:dos_pure} illustrates the local learnability of LDOS for pure silicon and pure carbon structures. 


 \begin{figure*}
 \includegraphics[width=.20\textwidth]{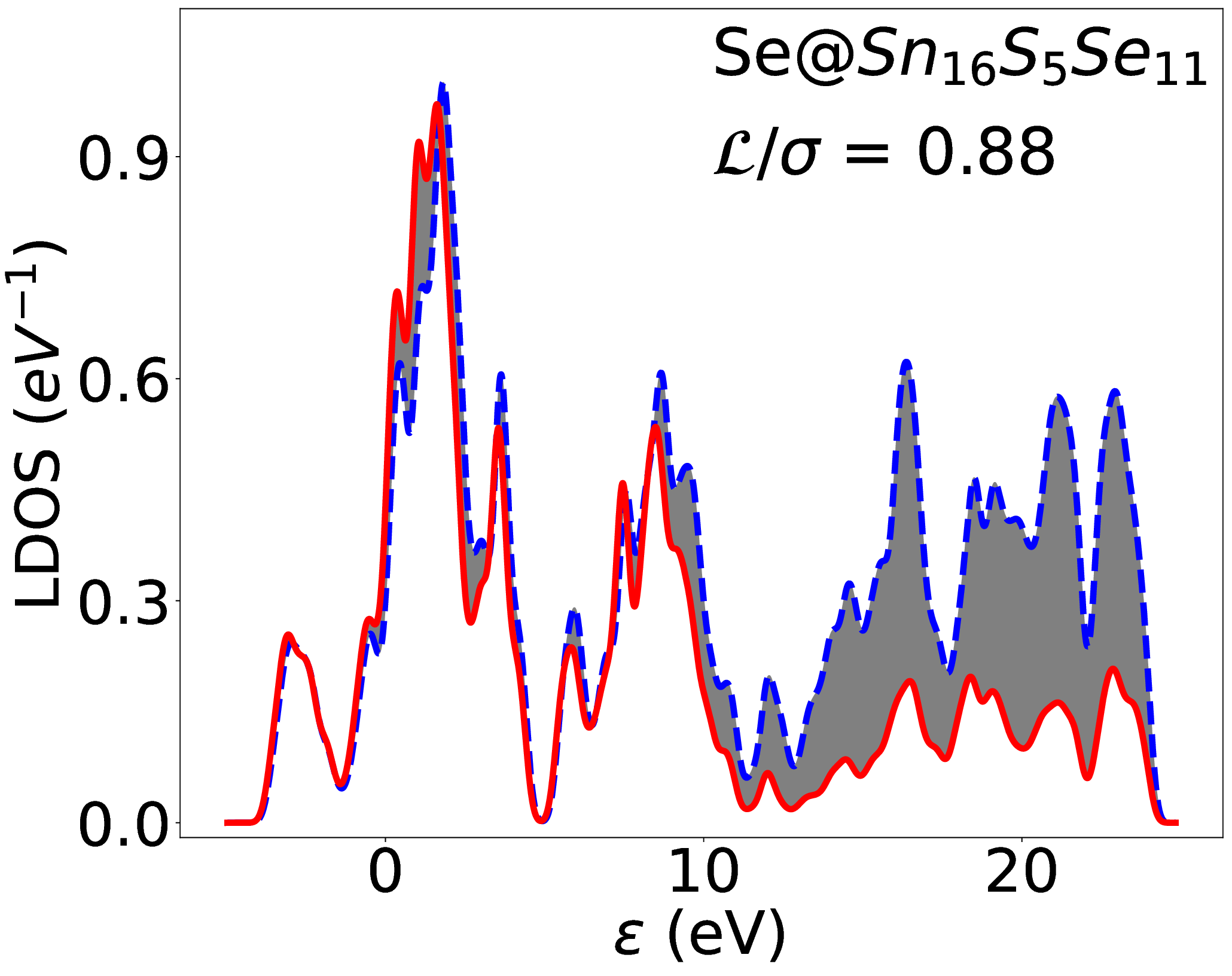}
 \includegraphics[width=.19\textwidth]{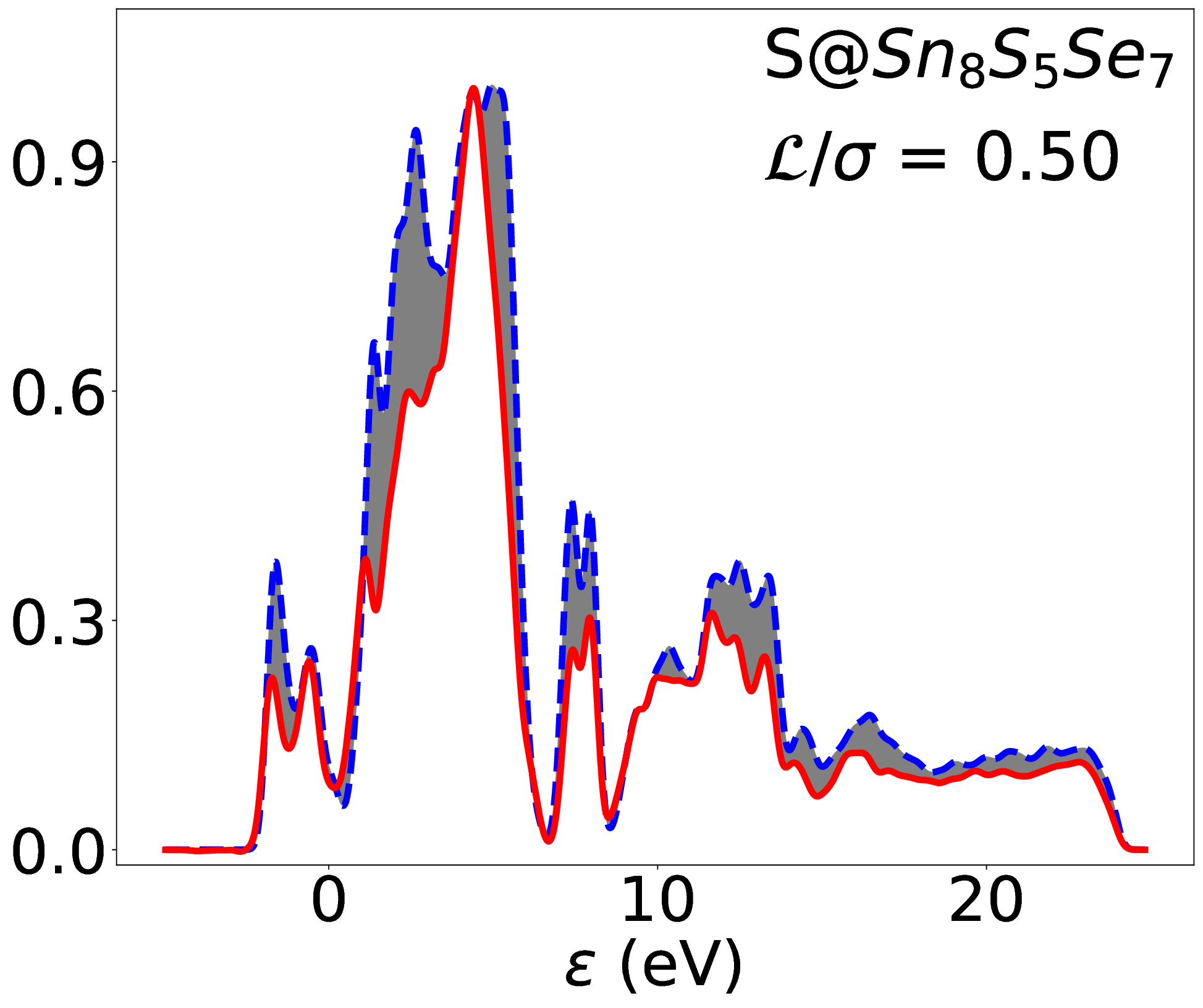}    
 \includegraphics[width=.19\textwidth]{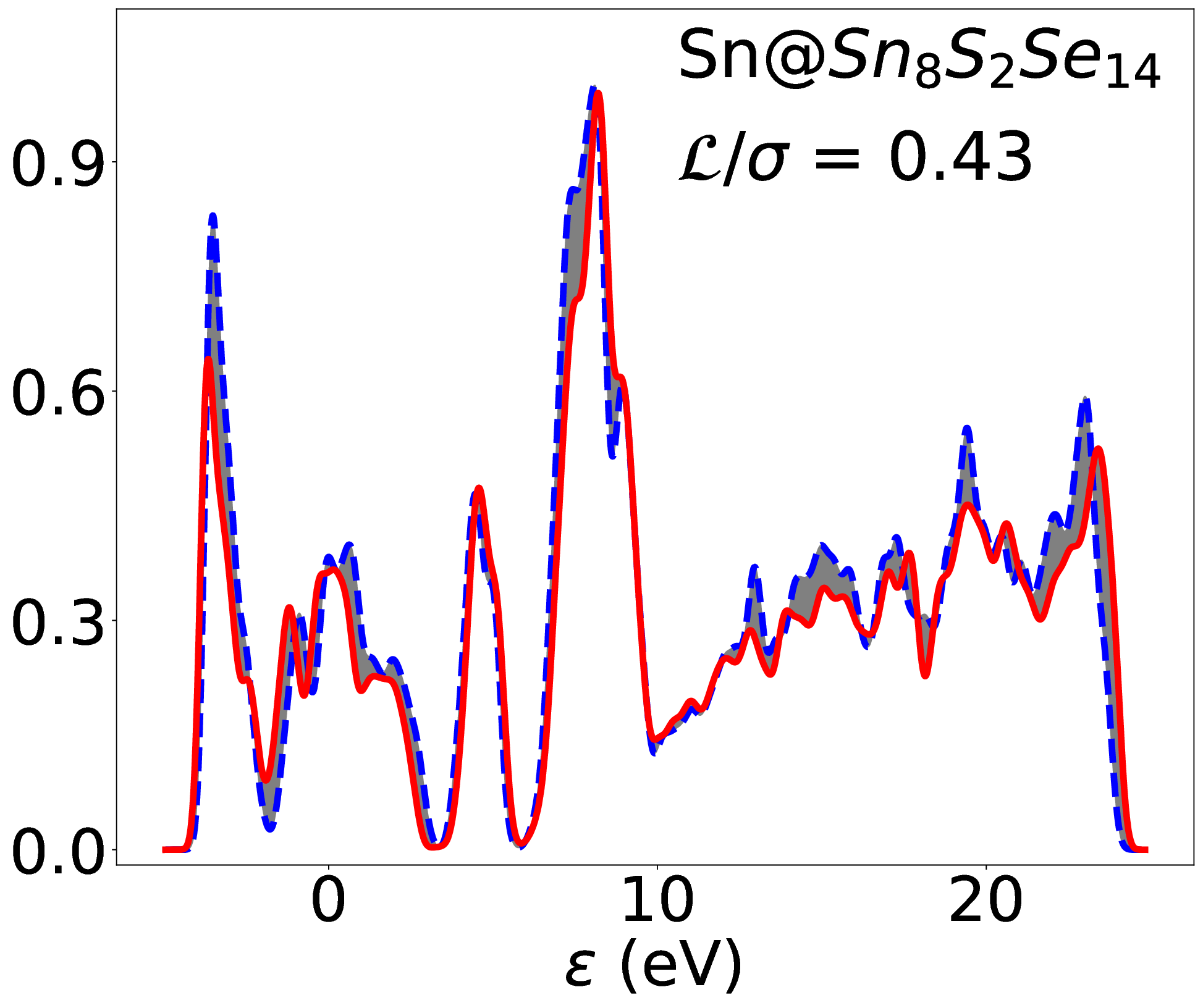}
 \includegraphics[width=.19\textwidth]{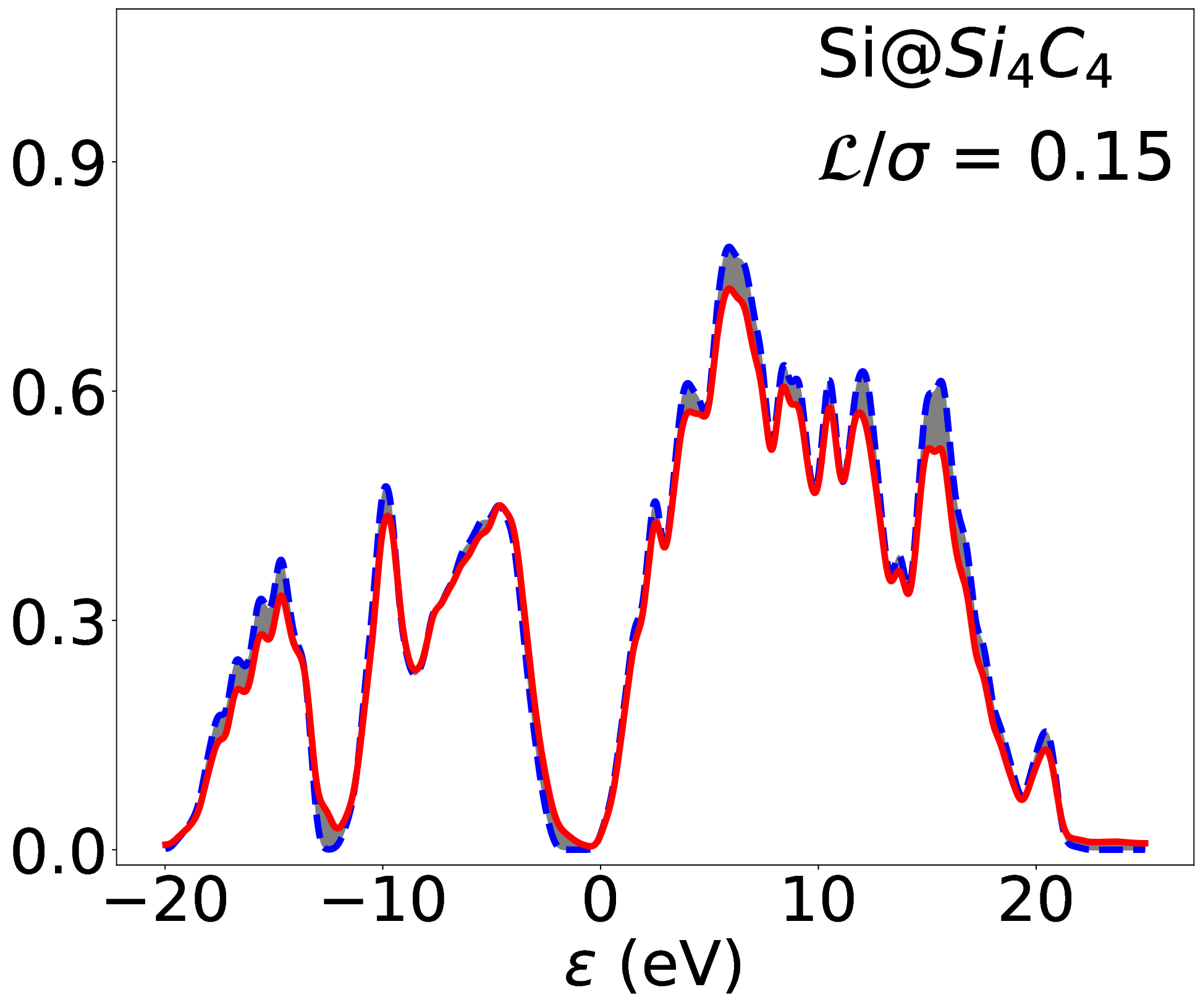}
 \includegraphics[width=.19\textwidth]{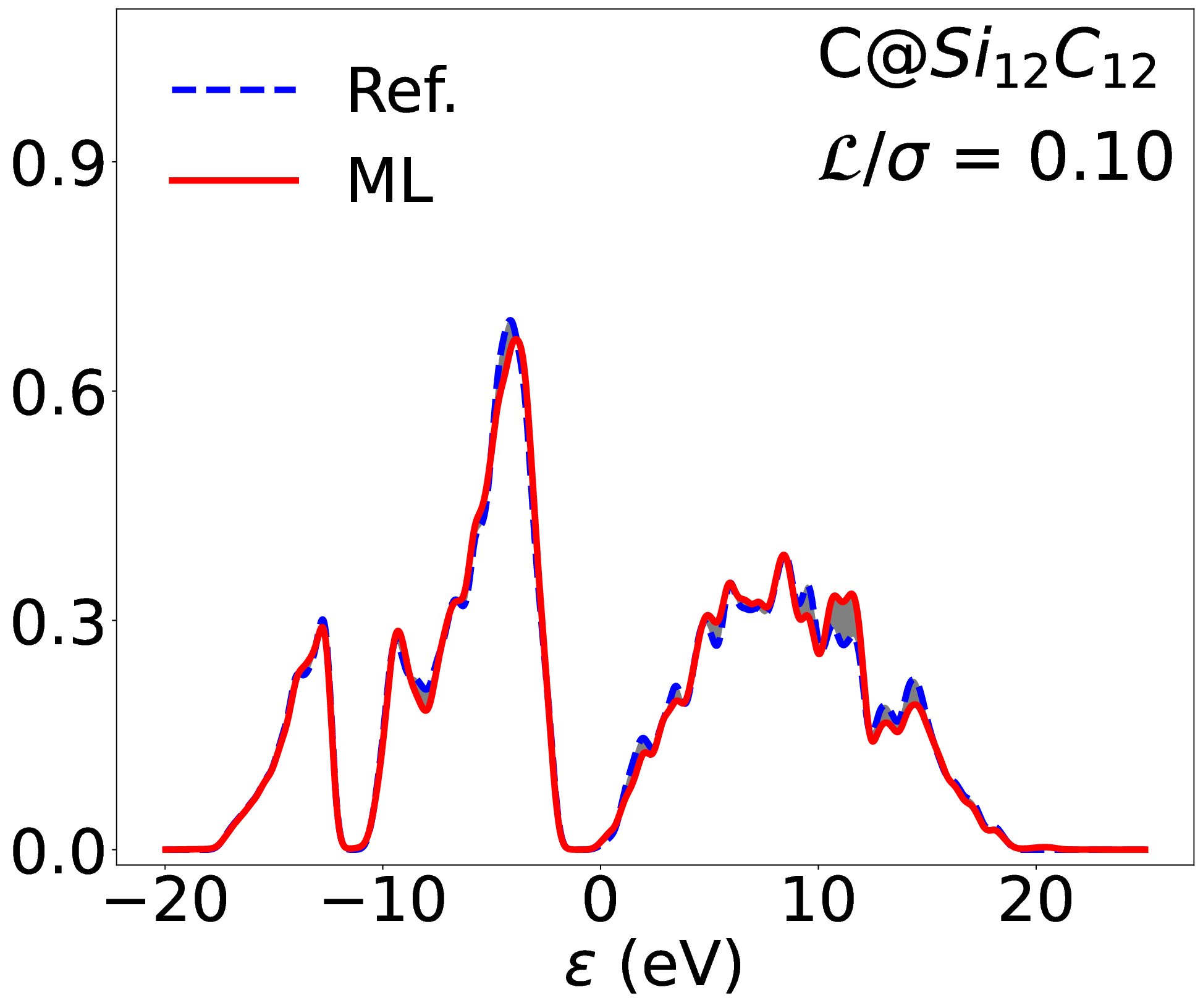}
\\
\includegraphics[width=.20\textwidth]{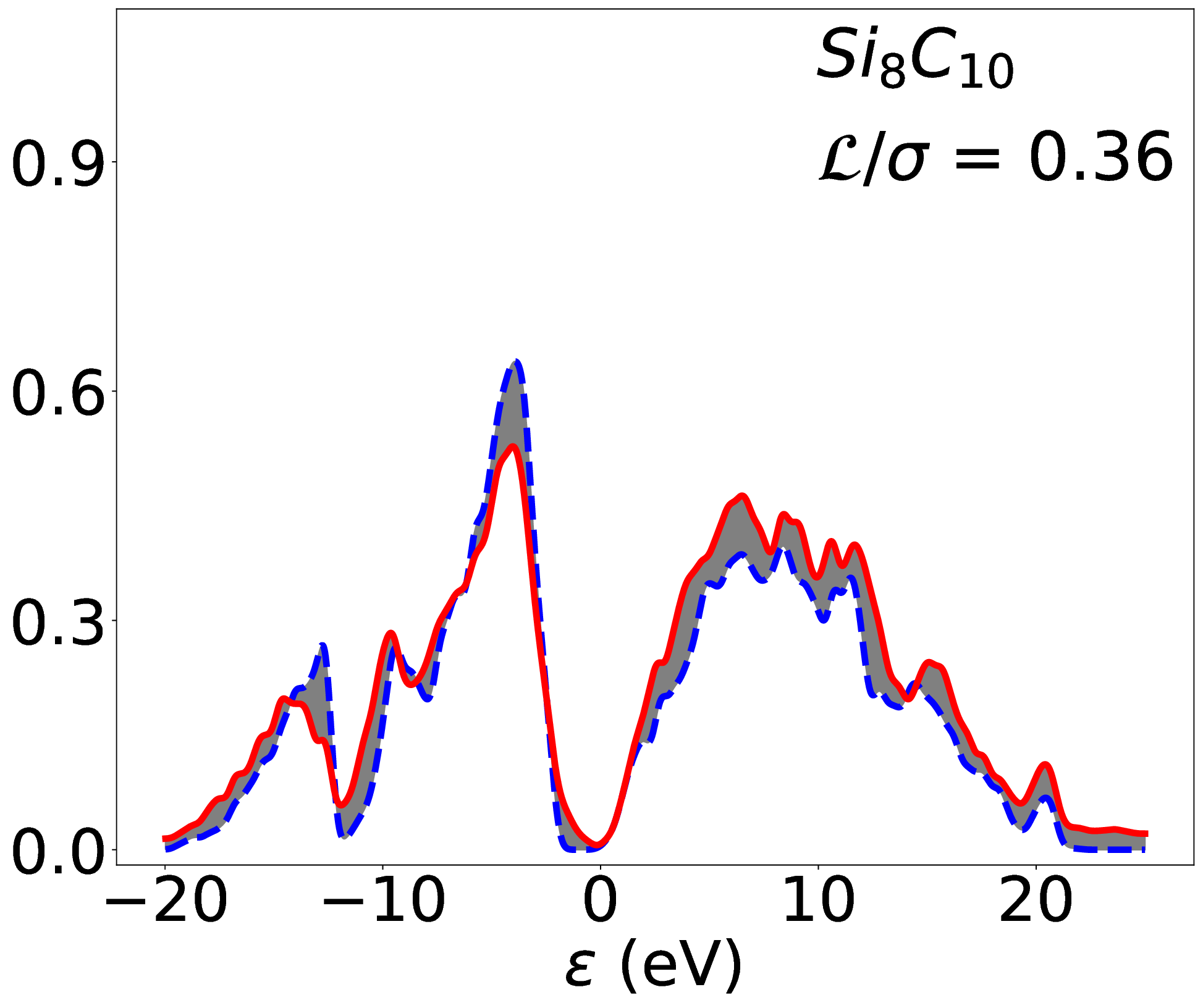}
\includegraphics[width=.19\textwidth]{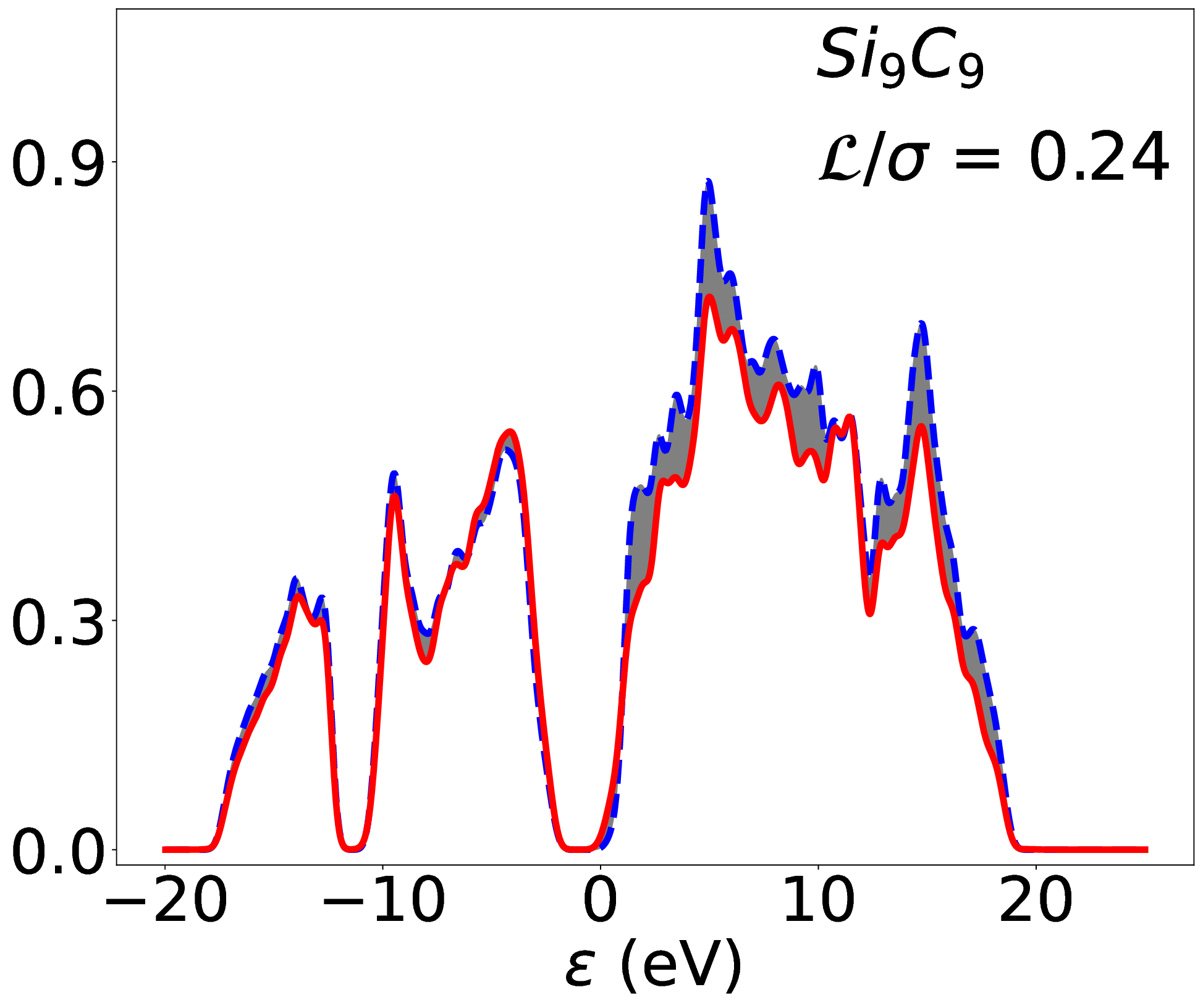}
\includegraphics[width=.19\textwidth]{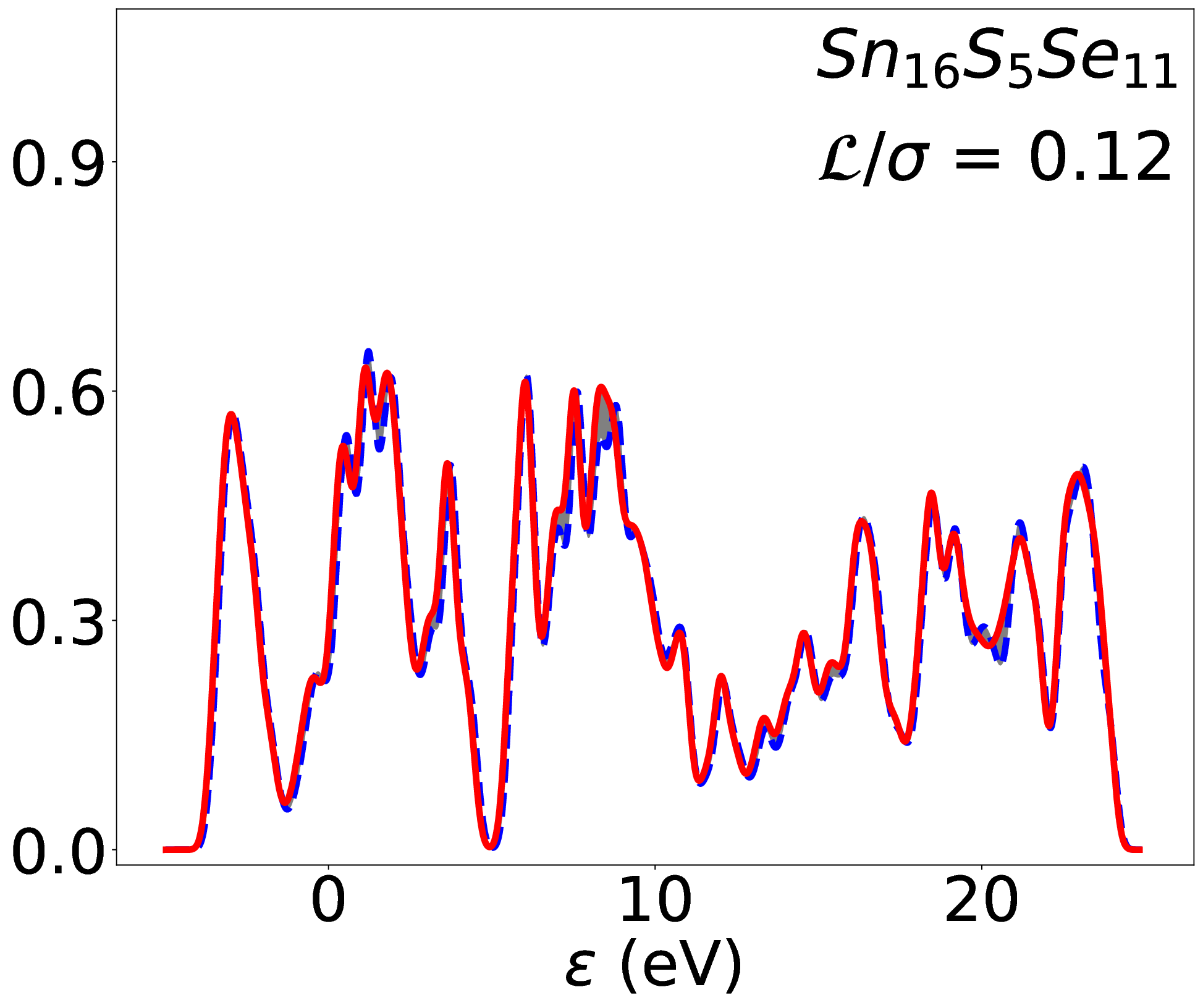}
\includegraphics[width=.19\textwidth]{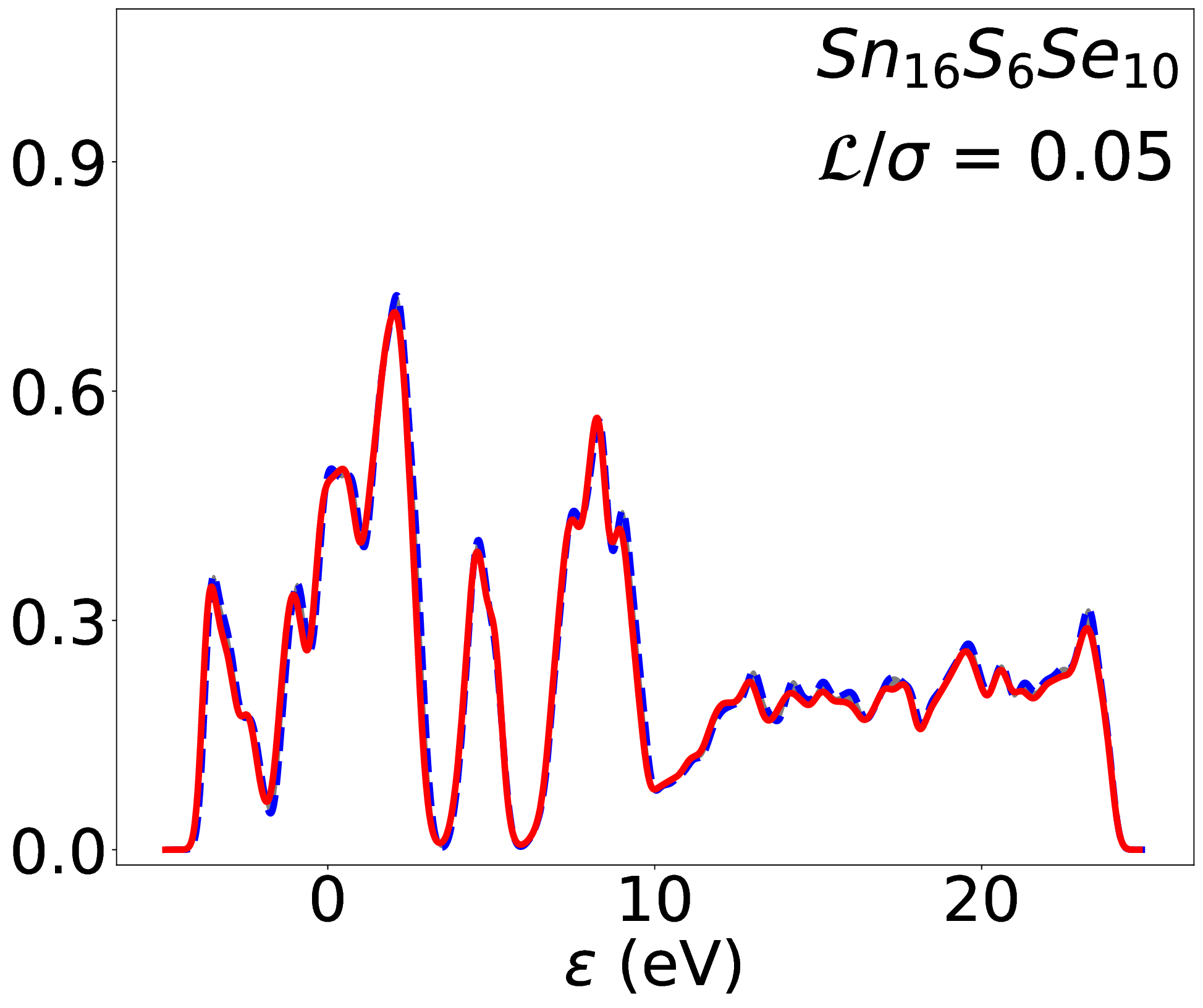}
\includegraphics[width=.19\textwidth]{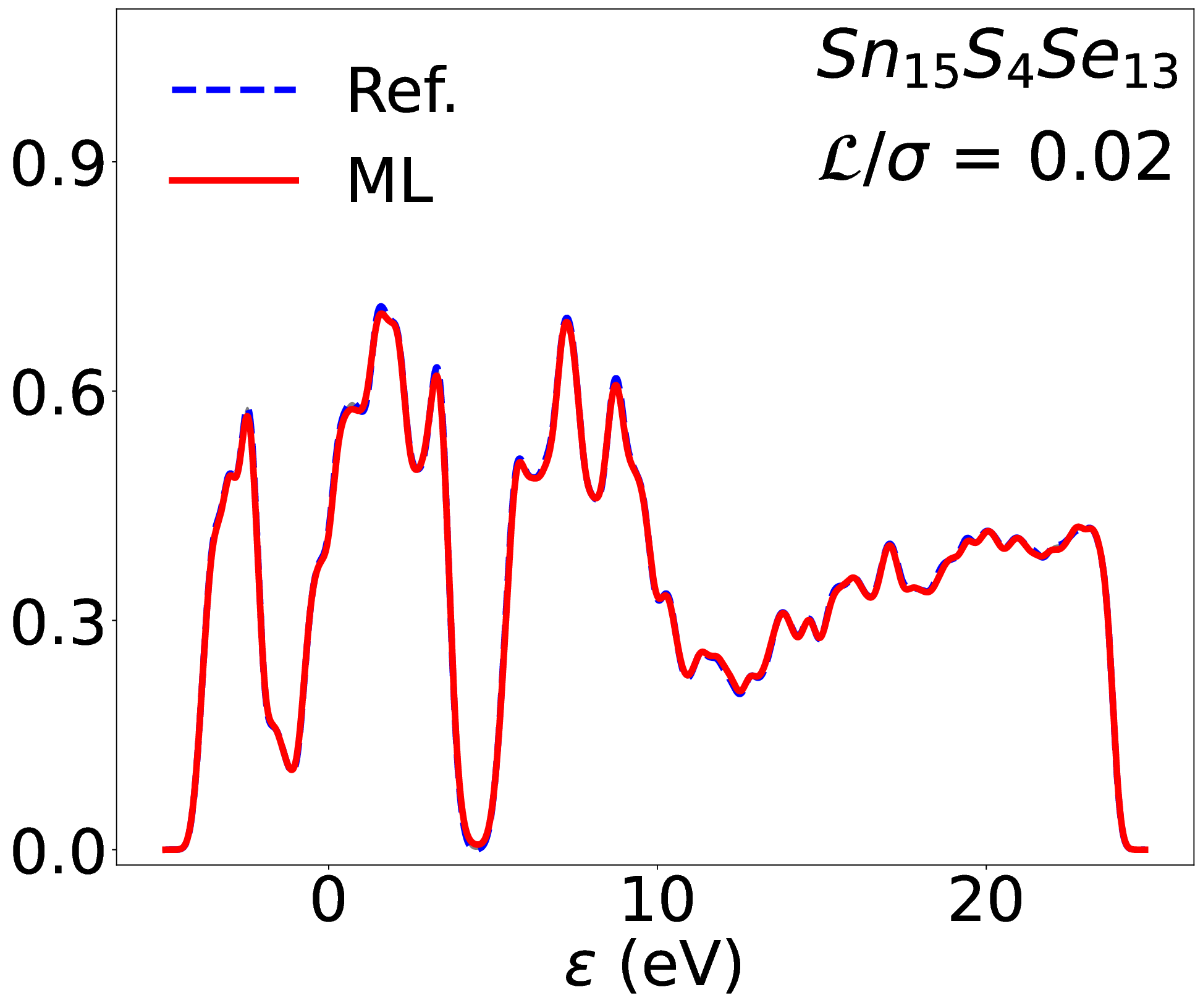}
   
\caption{\label{fig:dos_compound}  
	   Same as Fig.~\ref{fig:dos_pure} but for compound structures of C-Si (Set C) or Sn-S-Se (Set D).
  for two typical silicon (top row) and two typical carbon (bottom row) atoms.
	 Each row includes the worst prediction for every element
	 in terms of the normalized loss $\mathcal{L}/\sigma$. 
	 The bottom row also includes the best predictions. 
  }
  \end{figure*}

 \emph{Compound structures.} 
  We next consider chemically more complicated structures containing two or three alike atoms.
 To this aim, we use the compound Si-C structures from Set C and the Sn-S-Se structures  from Set D.
  The predicted LDOS and DOS are compared with the reference ones in Fig.~\ref{fig:dos_compound}
  for five representative cases.
  Note that the representatives for the LDOS include only the worst case for all five elements.
  Notably, the prediction error for the LDOS of carbon or silicon atoms in the carbon-silicon compound,
  $\mathcal{L} \lesssim 0.25~\sigma$, 
  remains in the same range as the pure structures.
  The predicted DOS, however, is slightly larger (0.36~$\sigma$)
  but is still comparable to those of Fig.~\ref{fig:dos_pure}.
 
  In the Si-C compound, all bonds are covalent and rather
  similar to the pure carbon and pure silicon structures.
This can be the reason why the accuracy of predicting the LDOS for 
Si and C atoms in the C-Si is comparable to the pure structures.
  In contrast, different interatomic bonds exist in the Sn-S-Se structures
which decreases the accuracy of learned LDOS in this compound. 
The worst observed case is a Se atom in a Sn$_{16}$S$_{5}$Se$_{11}$ cell with $\mathcal{L}/\sigma = 0.88$.
However, inspecting the LDOS profile reveals that the deviation from the reference is 
mainly for the states with energies above the Fermi energy.
The worst prediction of LDOS for the S atom occurs for a 
Sn$_8$S$_5$Se$_7$ cell with a loss of 0.5~$\sigma$  while for Sn the loss is at most 0.43~$\sigma$.
  Nevertheless, 
  the structural DOS of the three elements can be efficiently learned with a loss within $0.25~\sigma$,
as shown in the bottom row of Fig.~\ref{fig:dos_compound}.

\begin{table}[h]
\caption{\label{tab:test} 
Prediction error of atomic LDOS for simple and complex structures,
to verify model accuracy in terms of: 
  (i) scalability from small to large pure silicon structures,
 (ii) transferability between pure silicon,  pure carbon, and their binary compound, 
and (iii) generality to the complex solid solution Sn-S-Se. 
The training set excludes environments from the test structures.
} 
\begin{tabular} {l c c c c}
 & & \multicolumn{3}{c}{Error ($\mathcal{L}/\sigma$)}  \\ 
\cline{3-5} Atomic Species \; & \# Env. \;  & Min \,  &   Max \,  & {\bf Mean} \\ [0.5ex] \hline 
{Scalability:}  \\ 
Si  & 5616    & 0.030 & 0.17   & {\bf 0.09}  \\
	\hline
{Transferability:  }  \\ 
    Si   &  60  & 0.03  & 0.58   & {\bf 0.16}  \\
    C    &  68  & 0.18  & 0.52   & {\bf 0.31}  \\
 total   & 128  & 0.03  & 0.58   & {\bf 0.24}  \\
	\hline                            
{Generality:}  \\ 
   Sn   &  2400  & 0.026  & 1.54   & {\bf 0.24} \\
   S    &  1439  & 0.025  & 1.40   & {\bf 0.17} \\
   Se   &  1561  & 0.042  & 1.58   & {\bf 0.42} \\
total   &  5400  & 0.025  & 1.58   & {\bf 0.27} \\
 \hline 
\end{tabular}
\end{table}

Table~\ref{tab:test} summarizes the LDOS prediction error for three tests over a large number of test cases.
The first test, aimed at evaluating the model performance on pure structures,
demonstrates that the model  trained on small silicon structures in Set A
predicts the atomic LDOS of larger silicon structures in Set B 
     with a loss of $\mathcal{L} \leq 0.17~\sigma$.
     The average of the prediction error is only 9\% of $\sigma$.
The next test examines the model transferability between covalent structures of Set C.
When the train and test sets both consist of a collection of pure silicon, pure carbon and  silicon-carbon compound structures,
 the mean loss increases to 24\% of $\sigma$, while the maximum loss reaches $0.58~\sigma$.
Finally, in the third test on complex (solid solution) Sn-S-Se  compounds
the loss has an average  of $\mathcal{L}$ of $0.27~\sigma$,
   while its maximum  reaches  $\sim 1.58~\sigma$.
  The mean loss remains within $0.42~\sigma$ across all atomic species,
  with the selenium atoms exhibiting the least accuracy.
These preliminary tests suggest that  
  it is possible to train the LDOS model on mixed atomic types in compound   structures with heterogeneous interatomic bonds, although it is less efficient compared to simple covalent structures. 
A subject of future work is to investigate how  
adjusting the energy reference separately for individual atoms~\cite{adaptivedos25} 
improves the accuracy of predicting the LDOS in such complex structures. 
 Overall, the atomic local environment descriptor that is used,
	along with the employed machine learning model,
   successfully captures the varied bonding characteristics in the compounds.
This generalizability feature is practically significant,
 highlighting the potential of our model for application to complex atomic systems and compounds.

\section{Conclusions}\label{sec:conclusions}
  Electronic DOS is a global quantity that depends on the whole structure.
  We have showed that the DOS can be effectively decomposable into additive contributions known as atomic LDOS,
  which are assigned to and learned from the local environments of individual atoms.
  The prediction accuracy of LDOS significantly depends on the 
  size of the cropped spherical basin around each atom, as this basin contains the information essential for learning the atomic LDOS. 
  However, increasing the cropping  cutoff radius beyond 4-6~\AA, diminishes the size-dependence of to the prediction accuracy. 
  This behaviour implies that  atomic LDOS is a local quantity,
  and adheres to the Kohn's nearsightedness principle of electronic matter. 
Thanks to the locally learnable nature of LDOS,
  prediction of the structural DOS 
  is transferable between different geometries and compositions,
  while its additivity  ensures that the model is scalable and applicable to large structures. 
  Furthermore, LDOS provides a deeper physical understanding essential for accurate predicting properties that are significantly impacted by local electronic changes. 
  
  By comparing the accuracy of ML models in  predicting structural DOS and atomic LDOS, 
  we found that a ML model trained on LDOS outperforms the one trained on structural DOS. 
  Lower prediction errors are observed when predicting the structural DOS
  and various DOS-derived physical quantities such as Fermi energy, band energy, DOS at the Fermi level, 
  photon adsorption spectrum and heat capacity.
  This improved accuracy arises because the  derivative of the DOS with respect to energy 
  is learned more accurately  when the machine learns LDOS instead of the structural DOS. 
  Finally, we have showed that the model is applicable to structures with complex interatomic bonds.
  Although the accuracy may not reach the level achieved  for pure structures, 
  our results indicate that atomic DOS can be effectively learned for different atomic species in a complex compound.

  \section*{Acknowledgments}
  AA would like to express his sincere gratitude to C. Ben Mahmoud for his invaluable technical guidance and insightful discussions. 
  Computational resources are provided by the SARMAD cluster.

  \appendix
  \section {Expressions of Physical Quantities in Terms of DOS} \label{app:formula}
  DOS is a crucial concept in solid-state physics, representing the number of states available at a given energy level for electrons in a system. 
  In the following, we review the relation between the physical quantities explored in the main text
  with the structural DOS.

At zero temperature, the chemical potential of a system coincides with the Fermi energy,   
$\mu = \varepsilon_F$,  such that   
$\int_{-\infty}^{\varepsilon_F} \mathcal{D}(\varepsilon) \, d\varepsilon $ represents the total number of electrons. 
At finite temperature, however, we must impose the condition  
\begin{equation}\label{eq:eF}  
    \int_{-\infty}^{\infty} \mathcal{D}(\varepsilon) f_\text{FD}(\varepsilon) \, d\varepsilon = \text{No. of electrons},  
\end{equation}  
by adjusting the chemical potential in the Fermi-Dirac distribution
\begin{equation}\label{eq:FD}  
    f_\text{FD}(\varepsilon) = \left[1 + \exp\left(\frac{\varepsilon - \mu(T)}{k_B T}\right)\right]^{-1}.  
\end{equation}  
Consequently, the DOS at the Fermi energy, \(\mathcal{D}(\varepsilon_F)\), can be determined by interpolating the known values of the DOS around \(\varepsilon_F\).   
Similarly, the electronic contribution to the band energy is expressed as   
\begin{equation}\label{eq:Eb}  
    \varepsilon_{B} = \int_{-\infty}^{\infty} \varepsilon \mathcal{D}(\varepsilon) f_\text{FD}(\varepsilon) \, d\varepsilon,  
\end{equation}  
which facilitates the calculation of several interesting derived quantities.  
For example, 
the electronic contribution to the heat capacity
is expressed in terms of the temperature derivative of Eq.~(\ref{eq:FD}), namely
\begin{equation}\label{eq:C}
	  C(T) =  \frac{\partial \varepsilon_b}{\partial T} 
   = \int_{-\infty}^{\infty} 
  \varepsilon\mathcal{D}(\varepsilon)
  \left(\frac{\partial{f_\text{FD}}} 
  {\partial T} \right) 
   \, d\varepsilon
. \end{equation}

  Magnetic susceptibility of a material   describes its magnetic response to an external magnetic field as
  $\chi = {\partial M}/{\partial B}$.
  For free electrons, the so-called Pauli paramagnetism due to the orientation of the spin magnetic moment of electrons
  can be formulated as follows. 
  The total magnetic moment 
  \begin{equation}\label{eq:M0} M = \mu_B (N_+ - N_-),\end{equation}
  where $\mu_B $ is the Bohr magneton,
  is the consequence of an unbalanced number of electrons at the spin-up and spin-down states, 
  $N_+$ and $N_-$, respectively.
  In the absence of magnetism,
  \begin{equation}\label{eq:Npm0} 
  N_+ = N_- = \frac 1 2
  \int_{-\infty}^{\infty} \mathcal{D}(\varepsilon) f_\text{FD}(\varepsilon) \, d\varepsilon,
  \end{equation}
  where $f_\text{FD}(\varepsilon)$ denotes the Fermi-Dirac distribution function; see Eq.~(\ref{eq:eF}).
  Applying an external magnetic field $B$, however,
  shifts down or up the energy of the electrons 
  by $ \pm \mu_B B$
  depending on their spin orientation with respect to the magnetic field direction.
  The energy shift causes that more number of the states with spin in the same direction as $B$ become occupied while 
  the occupancy of the antiparallel states decreases,
  so that Eq.~(\ref{eq:Npm0}) is replaced by   
  \begin{multline}\label{eq:Npm} 
  N_\pm= \frac 1 2
  \int 
	  \mathcal{D}(\varepsilon \pm  \mu_B B) f_\text{FD}(\varepsilon) \, d\varepsilon 
  \\
  = \frac 1 2 \int \mathcal{D}(\varepsilon) f_\text{FD}(\varepsilon\mp  \mu_B B) \, d\varepsilon 
  .\end{multline}
  Then, the magnetic moment, Eq.~(\ref{eq:M0}),  reads
  \begin{multline}\label{eq:M} 
  M=\frac 1 2  \mu_B \int 
	  \mathcal{D}(\varepsilon) 
  \Big(f_\text{FD}(\varepsilon -\mu_B B)- f_\text{FD}(\varepsilon +\mu_B B )\Big) \, d\varepsilon 
   \\ 
  =      \mu_B^2 B
  \int 
	  \mathcal{D}(\varepsilon) 
  \left(-\frac{\partial f_\text{FD}}{\partial \varepsilon} \right) \, d\varepsilon 
  ,\end{multline}
  where we used a center finite difference approximation for small $B$ as
  $$
  \frac{\partial f_\text{FD}}{\partial \varepsilon}
  \simeq 
  \frac {f_\text{FD}(\varepsilon + \mu_B B)- f_\text{FD}(\varepsilon - \mu_B B) }
  {2\mu_B B} 
  .$$
  Finally, we obtain the magnetic susceptibility as
  \begin{align}
  \chi = \frac{\partial M}{\partial B} = \mu^2_B \int \mathcal{D}(\varepsilon)
  \left(-\frac{\partial f_\text{FD}}{\partial \varepsilon}\right)
  \,d\varepsilon 
  \label{eq:chi}
  .\end{align} 

Moreover, the adsorption and emission  spectrum of a structure is connected to its DOS. 
The probability of adsorption of a photon with energy $\hbar \omega$ depends on the number of available 
initially filled and initially empty electronic states that have an energy difference of $\hbar \omega$, 
namely~\cite{benmahmoud}
\begin{multline}  
\mathcal{A}(\omega) = \iint 
\mathcal{D}(\varepsilon) f_\text{FD}(\varepsilon)   
\mathcal{D}(\varepsilon') \left(1 - f_\text{FD}(\varepsilon')\right)           \\
\times \delta(\varepsilon - \varepsilon' - \hbar \omega) \, d\varepsilon \, d\varepsilon'.  
\end{multline}


%
\end{document}